\documentclass[prb,twocolumn,showpacs,amsmath,amssymb,superscriptaddress,bibnotes]{revtex4-2}
\usepackage{graphicx}
\usepackage{color}
\usepackage{braket}
\usepackage{amsmath}
\usepackage{amsfonts}

\usepackage{amsmath,amssymb,mathrsfs}
\usepackage{bbm}
\usepackage[dvipsnames]{xcolor}
\usepackage{appendix}
\usepackage{verbatim}
\usepackage{bbold}

\usepackage[unicode]{hyperref} 
\usepackage{blkarray}

\usepackage{soul}

\date{\today}

\begin{document}
	\title{Fractional chiral second-order topological insulator from a three-dimensional \\ array of coupled wires }

	\author{Viktoriia Pinchenkova}
	\affiliation{Department of Physics, University of Basel, Klingelbergstrasse 82, CH-4056 Basel, Switzerland}
	\author{Katharina Laubscher}
	\affiliation{Condensed Matter Theory Center and Joint Quantum Institute, Department of Physics, University of Maryland, College Park, MD 20742, USA}
	\author{Jelena Klinovaja}
	\affiliation{Department of Physics, University of Basel, Klingelbergstrasse 82, CH-4056 Basel, Switzerland}
	
	\date{\today}
	
\begin{abstract}
We construct a model of a three-dimensional chiral second-order topological insulator (SOTI) from an array of weakly coupled nanowires. We show that, in a suitable parameter regime, the interplay between rotating magnetic fields and spatially modulated interwire tunnelings leads to the opening of gaps in the bulk and surface spectrum of the system, while one or more chiral hinge states propagating along a closed path of one-dimensional hinges are left gapless. The exact path of these hinge states is determined by the hierarchy of interwire couplings and the boundary termination of the sample. Depending on the ratio between the period of the rotating magnetic field and the Fermi wavelength, our model can realize both integer and fractional chiral SOTIs. The fractional regime emerges in the presence of strong electron-electron interactions and features hinge states carrying only a fraction $e/p$ of the electronic charge $e$ for a positive odd integer $p$.
\end{abstract}
	
\maketitle

\section{Introduction}\label{sec:Introduction}

For conventional topological insulators (TIs) and superconductors, the so-called bulk-boundary correspondence predicts that a system with a topologically nontrivial $d$-dimensional bulk hosts topologically protected gapless states at its $(d-1)$-dimensional boundaries. Recently, however, the classification of topological systems has been further extended to also include so-called {\it higher-order} topological phases of matter~\cite{Benalcazar2017,Benalcazar2017b,Imhof2018,Song2017,Peng2017,Schindler2018,Geier2018,Langbehn2017}. In contrast to conventional TIs, $n$th-order TIs host protected gapless states only at their $(d-n)$-dimensional boundaries, while all higher-dimensional boundaries as well as the bulk remain insulating. For example, two-dimensional (three-dimensional) second-order TIs have a gapped bulk and gapped edges (gapped surfaces), but host topologically protected zero-energy corner states (gapless hinge states).

While the original theory of higher-order TIs (HOTIs) is based on single-particle band theory, it is also possible to extend some of the underlying concepts to strongly interacting systems~\cite{You2018,You2019,Laubscher2019,Laubscher2020,May-Mann2022,Hackenbroich2021,Zhang2022,Zhang2022b,Li2022,Laubscher2023}. This is particularly interesting since interaction-driven first-order phases such as the fractional quantum Hall (FQH) states are well-known to exhibit various exotic properties that cannot be realized in free-fermion systems. For example, quasiparticle excitations in these systems can carry fractional quantum numbers (e.g., fractional charge) and obey anyonic braiding statistics, making them potentially useful for topological quantum computing. It is therefore interesting to study if and how similar exotic phases---potentially in even richer variety---might emerge in the higher-order case. However, while non-interacting HOTIs are fairly well understood by now, the study of strongly interacting HOTIs is still at an early stage.

One of the challenges in this context is to study interaction-driven phases analytically at the microscopic level since electron-electron interactions have to be included nonperturbatively. Among the few approaches that allow one to construct analytically tractable toy models for strongly interacting phases is the so-called {\it coupled-wires} approach~\cite{Kane2002,Kane2014}, where higher-dimensional systems are constructed from arrays of weakly coupled one-dimensional (1D) wires. Within each wire, the electron-electron interactions can then be readily incorporated via standard 1D bosonization techniques~\cite{Giamarchi}, while the coupling between neighboring wires is taken into account as a small perturbation in a second step. This approach has turned out to be extremely useful in studies of various exotic interacting first-order topological phases of matter in two and three dimensions, including, for example, FQH states~\cite{Kane2002,Kane2014,Klinovaja2014b,Sagi2015b,Tam2021,Laubscher2021}, fractional quantum anomalous Hall (QAH) states~\cite{Klinovaja2015}, fractional TIs~\cite{Klinovaja2014,Sagi2014,Neupert2014,Santos2015,Sagi2015,Meng2015}, and fractional topological superconductors~\cite{Neupert2014,Sagi2017,Li2020}. Furthermore, it has been demonstrated that the coupled-wires approach can be applied to second-order topological phases, but only a very limited number of examples of such constructions exist to date~\cite{Laubscher2019,Laubscher2020,Zhang2022b,May-Mann2022,Laubscher2023}. 

In this work, we extend the family of known coupled-wires systems by constructing a three-dimensional (3D) model that is capable of realizing various chiral second-order TI (SOTI) phases. This model captures both integer SOTI phases with $l$ chiral hinge states (here $l$ is a positive integer) and fractional SOTI phases that emerge in the presence of strong electron-electron interactions. The latter feature hinge states that carry a fractional charge $e/p$, where $e$ is the elementary electron charge and $p$ is a positive odd integer. Furthermore, the topological phases constructed here can be seen as second-order QAH phases since the total magnetization in the model is set to zero. 

The paper is organized as follows. In Sec.~\ref{sec:Model}, we introduce the 3D coupled-wires model that we study in this work. In Sec.~\ref{sec:Integer}, we demonstrate that, for a suitable choice of system parameters, this model hosts gapless chiral hinge states that propagate along a closed path of hinges of a finite 3D sample. In Sec.~\ref{sec:Fractional}, we extend these considerations to the fractional case using bosonization techniques. We show that, for sufficiently strong electron-electron interactions, our model can realize a fractional chiral SOTI phase with gapless chiral hinge states carrying only a fraction of the electronic charge. Finally, we conclude in Sec.~\ref{sec:conclusion}.

\section{Model}\label{sec:Model}
\subsection{Main model based on helical magnetic fields}\label{subsec:main_model}

\begin{figure}[tb]
    \centering   
	\includegraphics[width=0.48\textwidth]{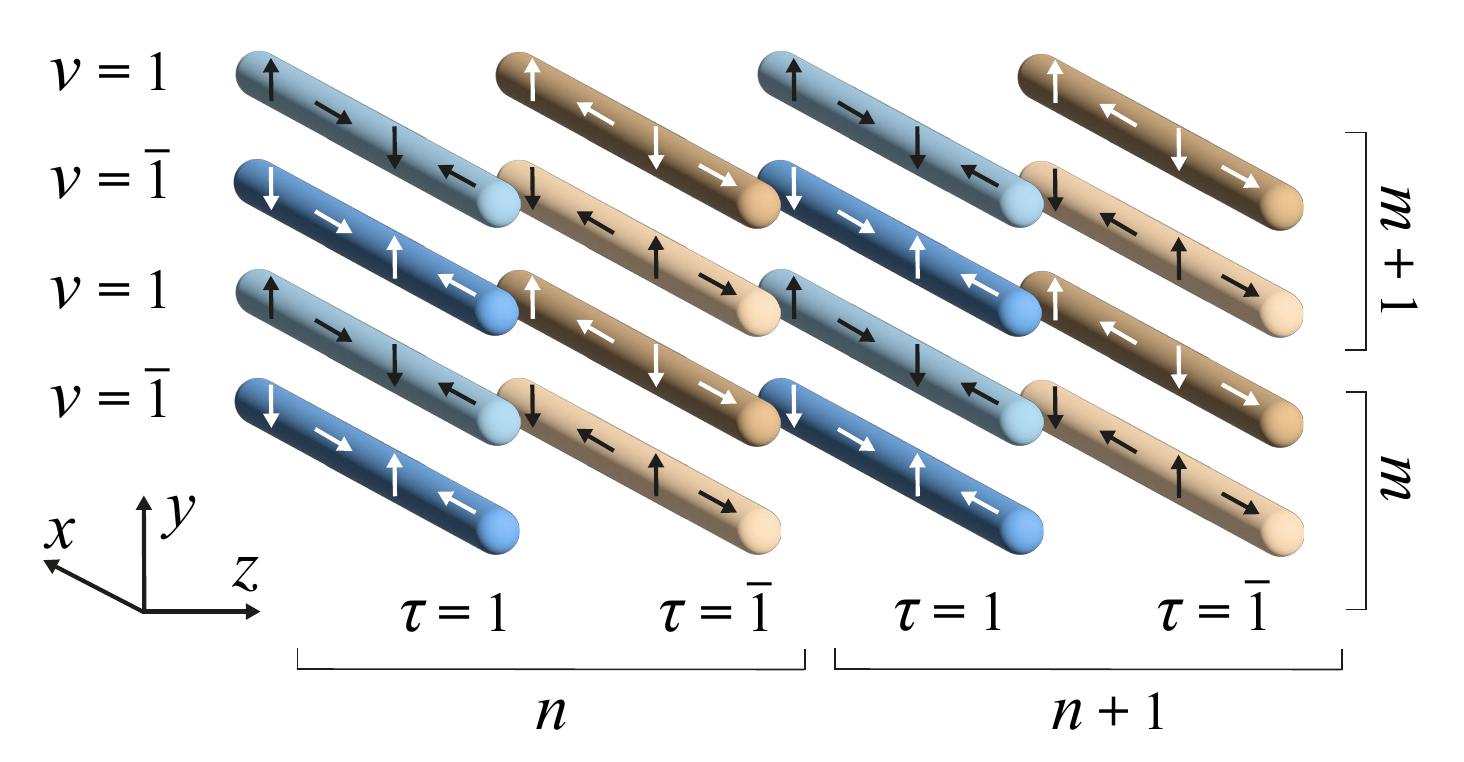}
    \caption{Sketch of the 3D construction composed of 1D nanowires aligned along the $x$ axis. The unit cell consists of four nanowires. Each wire is labeled by indices $(n, m, \tau, \nu)$, where the discrete indices $(n, m)$ denote the position of the unit cell, and $\tau \in \{1, \bar{1} \}$, $\nu \in \{1, \bar{1} \}$ denote the position of the wire inside the unit cell. The magnetic field acting inside each wire rotates clockwise (counterclockwise) in the $xy$ plane for wires with $\tau \nu = 1$ ($\tau \nu = \bar{1}$) as shown by black (white) arrows. In the coupled-wires approach, each wire is first considered independently and then the coupling between neighboring wires is introduced perturbatively.
	}
    \label{fig:model}
\end{figure}

We construct a model for a 3D SOTI from an array of weakly coupled 1D nanowires aligned along the $x$ direction, as shown in Fig.~\ref{fig:model}. A unit cell in our model is composed of four wires. The position of a unit cell is denoted by indices $n$ in the $z$ direction and $m$ in the $y$ direction, while within a unit cell, each wire is labeled by indices $\tau$ and $\nu$, with $\tau = 1$ ($\tau = \bar{1}$) denoting the left (right) wire relative to the $z$ axis, and $\nu = 1$ ($\nu = \bar{1}$) the top (bottom) wire relative to the $y$ axis (see Fig.~\ref{fig:model}). The spin quantization axis is set to be along the $z$ direction.

We assume that neighboring wires are weakly coupled with coupling amplitudes that are small compared to the chemical potential $\mu$ inside each wire. This assumption allows us to first treat each wire as completely independent and then add the coupling terms perturbatively. The kinetic part of the Hamiltonian describing the uncoupled wires is given by
\begin{equation}\label{eq:kinetic}
H_0 = \sum_{n, m} \sum_{\tau, \nu, \sigma} \int d x \, \Psi_{n m \tau \nu \sigma}^{\dagger}(x) \left[-\frac{ \partial_x^2}{2 m_0}-\mu \right] \Psi_{n m \tau \nu \sigma}(x),
\end{equation}
where $\Psi_{n m \tau \nu \sigma}^{\dagger}(x)$ and $ \Psi_{n m \tau \nu \sigma}(x)$ are the creation and annihilation operators of an electron with spin $\sigma \in \{1, \bar{1} \}$ at the position $x$ of the wire $(\tau, \nu)$ in the unit cell $(n,m)$. Furthermore, $m_0$ is the effective electron mass, and we put $\hbar = 1$. For infinitely long wires, the energy spectrum inside each wire takes a simple quadratic form: $E_{0} = k_x^2/2 m_0 - \mu$, where $E_0$ is twofold degenerate in spin $\sigma$. The Fermi wave vector $k_F$ is related to the chemical potential by $k_F = \sqrt{2 m_0 \mu}$.

Next, we introduce a magnetic field rotating in the $xy$ plane:
\begin{equation}\label{eq:magnetic_field}
\mathbf{B}_{\tau \nu} ^{(l)}(x)  = \tau B\left[\cos \left(2 k_F \frac{x}{l}\right) \hat{\mathbf{x}}-(\tau \nu ) \sin \left(2 k_F \frac{x}{l}\right) \hat{\mathbf{y}}\right], 
\end{equation}
where $\hat{\mathbf{x}}$ and $\hat{\mathbf{y}}$ are unit vectors in the $x$ and $y$ directions, and $B$ is the strength of the magnetic field. The dimensionless parameter $l$ determines the rotation period of the magnetic field, given by $\pi l /k_F$, and is assumed to be either a positive integer or a fraction of the form $l=1/p$ with $p$ being an odd positive integer. As we will see below, $l$ is related to the charge of the gapless hinge states in such a way that the total charge carried by these states is $e l$. The magnetic field $\mathbf{B}_{\tau \nu} ^{(l)}(x)$ rotates clockwise (counterclockwise) in wires with $\tau \nu = 1$ ($\tau \nu = \bar{1}$), such that no total magnetization is created. The corresponding Zeeman term is
\begin{align}\label{eq:Zeeman1}
    H_{B}^{(l)} = \sum_{n, m} & \sum_{\tau, \nu}  \sum_{\sigma, \sigma^{\prime}}  \int d x \, 
    \Psi_{n m \tau \nu \sigma}^{\dagger}(x)  \nonumber\\ &\quad\times  \left (  \beta \, \mathbf{B}_{\tau \nu}^{(l)}(x) \cdot \boldsymbol{\sigma}  \right )_{\sigma \sigma^{\prime}}  \Psi_{n m \tau  \nu \sigma^{\prime}}(x),
\end{align}	
where $\beta$ is the coupling constant between the magnetic field and the electron spin, and the vector $\boldsymbol{\sigma} = (\sigma_1, \sigma_2, \sigma_3)$ is composed of the Pauli matrices $\sigma_i$ representing the electron spin. The Zeeman term can be rewritten in the form
\begin{align}\label{eq:Zeeman2}
    H_{B}^{(l)} = & \Delta_B  \sum_{n, m} \sum_{\tau, \nu}  \tau \int d x \, \Psi_{n m  \tau \nu 1}^{\dagger}(x)    \nonumber\\ &\quad\times  \Psi_{n m \tau  \nu \bar{1}}(x) \exp \left[ i (\tau \nu)  \: 2 k_{F} \frac{x}{l} \right] + \text{H.c.}
\end{align}	
The energy scale $\Delta_B = B \beta$ is assumed to be small compared to the chemical potential $\mu$. Note that although the total magnetization is zero, the magnetic fields still act locally on the electron spins within each wire and thereby break the time-reversal symmetry of the system.

The required rotating magnetic field defined in Eq.~(\ref{eq:magnetic_field}) can be generated in several ways, for example, by extrinsic nanomagnets~\cite{Karmakar2011,Klinovaja2012,Kjaergaard2012,helical_graphene}, a magnetic skyrmion texture~\cite{skyrmion1,skyrmion2,skyrmion3,skyrmion4,skyrmion5}, or helical local magnetic moments formed via Ruderman-Kittel-Kasuya-Yosida interactions~\cite{RKKY1,RKKY3,RKKY4,RKKY5}. Another option is to use a combination of spin-orbit interaction (SOI) in Rashba nanowires and uniform magnetic fields instead of the rotating fields~\cite{Braunecker2010}; see Sec.~\ref{subsec:SOI}.

We now introduce tunneling processes between neighboring wires. The tunneling along the $y$ direction is assumed to be position and spin dependent, giving
\begin{align}\label{eq:Hy}
    H_y^{(l)} = \sum_{n,m} \sum_{\tau, \sigma} \tau & \int d x \,
    t^{(l)}_y(x) \Psi_{n m \tau 1 \sigma}^{\dagger}(x)  \nonumber\\ &\quad\times \Psi_{n [m + (1-\sigma)/2]  \tau \bar{1} \sigma}(x)+ \text{ H.c.},
\end{align}	
where the tunneling amplitude depends on the position as $t^{(l)}_y(x)= 2 t_y \cos \left(2 k_F x / l\right)$ with the period $\pi l/k_F$ equal to the period of the magnetic field $\mathbf{B}_{\tau \nu} ^{(l)}(x)$, and with $t_y$ being a non-negative constant. In general, the main results of the paper remain valid for other forms of tunneling amplitudes and magnetic field, including potentially more experimentally feasible forms, as long as they have a substantial Fourier component at $2 k_F/l$. The spin dependence of the tunneling processes can be realized by placing nanomagnets to polarize the medium between the wires or via intrinsic magnetic ordering. As a result, only spin-up (spin-down) electrons are allowed to tunnel between neighboring wires of the same unit cell (in adjacent unit cells); see Fig.~\ref{fig:spectrum_l=1}.

Finally, we add the tunneling processes along the $z$ direction. The first term couples neighboring wires in adjacent unit cells: 
\begin{equation}\label{eq:Hz}
H_z^{(l)} = \sum_{n, m} \sum_{\nu, \sigma} \int d x \, t_{z\nu}^{(l)}(x) \Psi_{(n+1) m 1 \nu \sigma}^{\dagger} \Psi_{n m \bar{1} \nu \sigma}+\text { H.c.},
\end{equation}
while the second term couples neighboring wires of the same unit cell:
\begin{equation}\label{eq:Hz'}
\widetilde{H}_{z}^{(l)} = \sum_{n, m} \sum_{\nu, \sigma} \int d x  \, \tilde{t}_{z \nu }^{(l) }(x) \Psi_{n m 1 \nu \sigma}^{\dagger} \Psi_{n m \bar{1} \nu \sigma}+\text { H.c.}
\end{equation}
Hereinafter we omit the position argument of the field operators for brevity. The magnitudes of the tunneling amplitudes are spatially modulated as $t_{z \nu }^{(l) }(x) = 2 t_{z \nu} \cos \left(2 k_F x/l\right)$ and $\tilde{t}_{z \nu }^{(l) }(x) = 2 \tilde{t}_{z \nu} \cos \left(2 k_F x /l\right)$, where $t_{z \nu}$ and $\tilde{t}_{z \nu}$ are non-negative constants depending on $\nu$.

The total Hamiltonian of our model $H^{(l)}$ is given by the sum of all terms described above:
\begin{equation}\label{eq:Htotal}
H^{(l)} = H_0 +H_{B}^{(l)} + H_y^{(l)} + H_z^{(l)} + \widetilde{H}_{z}^{(l)}.
\end{equation}
In Secs.~\ref{sec:Integer} and \ref{sec:Fractional}, we explicitly demonstrate that this model can realize various 3D SOTI phases with a fully gapped bulk and fully gapped surfaces, but gapless chiral hinge states with total charge $el$ that propagate along a closed path of hinges.

\subsection{Alternative model based on SOI}\label{subsec:SOI}

In the rest of the paper, we focus on the model with the Hamiltonian $H^{(l)}$ described in the previous subsection. However, as was mentioned above, this model is mathematically equivalent to an alternative model in which the helical magnetic fields are replaced by the interplay between SOI and uniform magnetic fields. We thus consider Rashba nanowires with strong SOI in this subsection. The Rashba SOI leads to an additional term in the Hamiltonian:
\begin{align}\label{eq:SOI}
    H_{\mathrm{SOI}} = - i \alpha \sum_{n, m} \sum_{\tau, \nu} \sum_{\sigma, \sigma^{\prime}} & \tau \nu  \int d x \,
    \Psi_{n m \tau \nu \sigma}^{\dagger}  \nonumber\\ &\quad\times \left( \sigma_3  \right)_{\sigma \sigma^{\prime}} \partial _x \Psi_{n m \tau  \nu \sigma^{\prime}},
\end{align}
where $\alpha$ parametrizes the strength of the SOI. This term sets the spin quantization axis along the $z$ direction and has a sign determined by $\tau \nu \in \{1, \bar{1} \}$. The kinetic part $H_0$, combined with the SOI term $H_{\mathrm{SOI}}$, results in the energy spectrum of infinitely long wires in the form:
\begin{equation}
E_{\tau \nu \sigma} = \frac{k^2_x}{2m_0} + (\tau \nu \sigma) \alpha k_x - \mu^{(l)}.
\end{equation}
Here, the chemical potential is tuned to the value $\mu^{(l)} = E_{so} (l^2 - 1)$, with $E_{so} = k_{so}^2/(2m_0)$ being the spin-orbit energy, and $k_{so} = m_0 \alpha$ the spin-orbit momentum. The new Fermi wave vectors satisfy $k_{F\pm}^{(l)} = k_{so}(1 \pm l)$.

Next, we introduce uniform magnetic fields with an amplitude $M$ applied along the wires. These magnetic fields are oriented in opposite directions for wires with different $\tau$ such that the total magnetization is equal to zero. The corresponding Zeeman term reads
\begin{equation}
H_{M} = \Delta_M \sum_{n, m} \sum_{\tau, \nu} \sum_{\sigma} \tau \int d x \, \Psi_{n m \tau \nu \sigma}^{\dagger} \Psi_{n m \tau  \nu \bar{\sigma}},
\end{equation}
where $\Delta_M = \beta M$. The tunneling terms remain the same as in the previous subsection~\ref{subsec:main_model}, but the tunneling amplitudes now have periods independent of $l$, specifically $\pi/(2 k_{so})$. The model described here is mathematically equivalent to the model based on rotating magnetic fields (see Sec.~\ref{subsec:main_model}) and can therefore also host integer and fractional chiral hinge states in the exact same way; for more details see Appendix~\ref{App:SOI}.

\section{Integer chiral hinge states}\label{sec:Integer}

\subsection{Single hinge state regime}\label{subsec:Single}

In this section, we demonstrate that the model defined in Eq.~(\ref{eq:Htotal}) realizes a SOTI phase in a certain region of parameter space. Here, we focus on the case of integer $l$, which leads to hinge states that carrying integer charge, while the case of fractionally charged hinge states will be discussed further below in Sec.~\ref{sec:Fractional}. For simplicity, we start with the single-state regime characterized by the parameter $l = 1$.

For now, let us assume that the system is infinite along the $x$ axis, but has a finite number of unit cells $N_y$ and $N_z$ in the $y$ and $z$ directions, respectively. To demonstrate that the system can realize a SOTI phase, we focus on the parameter regime $\mu \gg \Delta_B \gg t_y \gg t_{z 1}, \tilde{t}_{z \bar{1}}\gg  t_{z\bar{1}}, \tilde{t}_{z 1}  \geq 0$. In this limit, we can apply a multi-step perturbation procedure to show that the system hosts chiral hinge states propagating along the $x$ direction. As a first step, since all energy scales are small compared to the chemical potential $\mu$, we linearize the spectrum around the Fermi points $\pm k_F$ by rewriting the Hamiltonian in a basis composed of slowly varying right moving fields $R_{n m \tau \nu \sigma}(x)$ and left moving fields $L_{n m \tau \nu \sigma}(x)$ inside each wire (see, e.g., Ref.~\cite{Giamarchi}):
\begin{equation} 
\Psi_{n m \tau \nu \sigma} = e^{i k_F x} R_{n m \tau \nu \sigma} + e^{-i k_F x} L_{n m \tau \nu \sigma}. 
\end{equation}
In what follows, to simplify our notation, we work in terms of the Hamiltonian density $\mathcal{H}$ determined by $H=\sum_{n, m} \int d x \, \mathcal{H}(x)$ and present all Hamiltonian terms in this form. In terms of the new fields, the kinetic part corresponding to Eq.~(\ref{eq:kinetic}) becomes
\begin{equation}\label{eq:H0}
	\mathcal{H}_0 = -i v_F \sum_{\tau, \nu, \sigma}  (R_{n m \tau \nu \sigma}^{\dagger}  \partial_x R_{n m \tau \nu \sigma} -L_{n m \tau \nu \sigma}^{\dagger} \partial_x L_{n m \tau \nu \sigma} ),
\end{equation}
with the Fermi velocity $v_F = k_F/m_0$. Hereinafter, we neglect all fast-oscillating contributions.

\begin{figure}[tb]
	\centering   \includegraphics[width=0.47\textwidth]{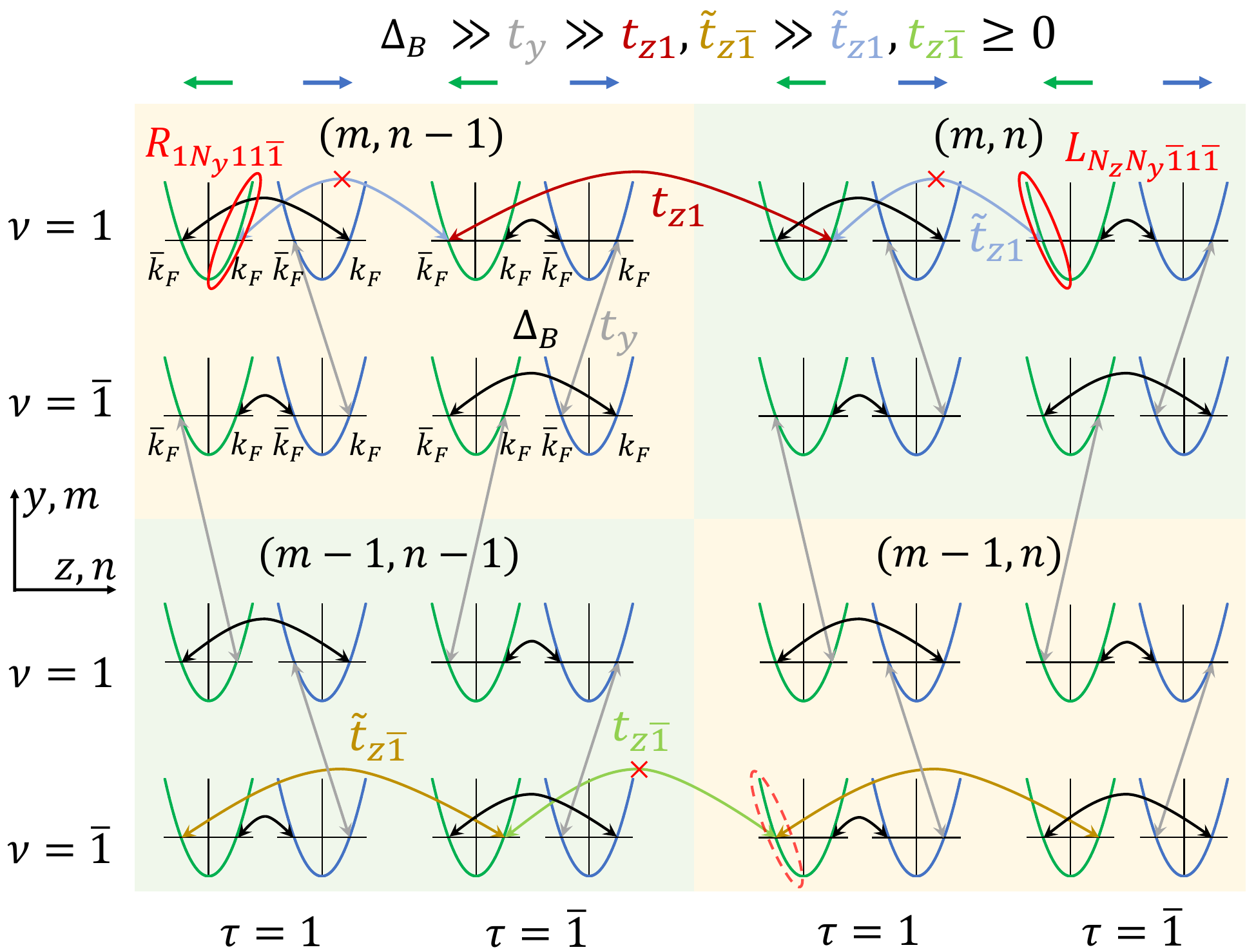}
	\caption{Sketch of four unit cells of the coupled-wires model with $l=1$. Parabolas represent the energy spectrum $E_0$ of electrons freely propagating along the wires. For clarity, initially twofold degenerate parabolas are shown separately: green (blue) parabolas represent spin-down (spin-up) electrons. The spin quantization axis is set along the $z$ axis, and the direction of spins is shown by green and blue arrows. The double arrows represent the magnetic and interwire terms in leading order of the perturbation theory. These terms result in a fully gapped bulk, fully gapped surfaces, and two gapless chiral states $R_{1 N_y 1 1 \bar{1}}(x)$, $L_{N_z N_y \bar{1} 1 \bar{1}}(x)$ localized to two hinges of the sample (denoted by red solid ovals). Changing the boundary termination by removing the last layer in the $xy$ plane at $n = N_z$, $\tau = \bar{1}$ leads to the relocation of one hinge state from $L_{N_z N_y \bar{1} 1 \bar{1}}(x)$ to $L_{N_z 1 1 \bar{1} \bar{1}}(x)$ (denoted by the red dashed oval).}
	\label{fig:spectrum_l=1}
\end{figure}

Next, we focus on the Zeeman and $y$-tunneling terms given by Eqs.~(\ref{eq:Zeeman2}) and (\ref{eq:Hy}) as these terms are assumed to be dominant in our parameter hierarchy. In terms of the right and left movers, these terms take the form
\begin{align}\label{eq:Zeeman_integer}
\mathcal{H}_B^{(1)} &= \Delta_B \sum_{\tau} \tau \left( \right.R_{n m \tau \tau 1}^{\dagger}  L_{n m \tau  \tau \bar{1}} \nonumber\\
&\hspace{18mm}+ L_{n m \tau \bar{\tau} 1}^{\dagger} R_{n m \tau \bar{\tau} \bar{1}} \left. \right) +\text { H.c.},
\end{align}
\begin{align}
\mathcal{H}_y^{(1)} &= t_y \sum_{\tau, \sigma} \tau (R_{n m \tau 1 \sigma}^{\dagger} L_{n [m + (1-\sigma)/2] \tau \bar{1} \sigma} \nonumber\\
&\hspace{13mm} + L_{n m \tau 1 \sigma}^{\dagger} R_{n [m + (1-\sigma)/2] \tau \bar{1} \sigma}) +\text { H.c.}
\end{align}
At this point, we see that the states $L_{n 1 1 \bar{1} \bar{1}}$, $R_{n N_y 1 1 \bar{1}}$, $R_{n 1 \bar{1} \bar{1} \bar{1}}$ and $L_{n N_y \bar{1} 1 \bar{1}}$ do not enter either one of these terms, and, hence, they remain gapless for now. These gapless states are located on the $xz$ surface of the sample. On the other hand, all other states in the bulk and on the $xy$ surfaces of the sample are fully gapped out, see also Fig.~\ref{fig:spectrum_l=1}.

In the next step of the perturbation procedure, we take into account the tunneling processes in the $z$ direction. In the same way as before, we rewrite Eqs.~(\ref{eq:Hz}) and (\ref{eq:Hz'}) in the basis of the new fields:
\begin{align}
\mathcal{H}_z^{(1)} &= \sum_{\nu, \sigma} t_{z \nu} \left( \right. R_{(n+1) m 1 \nu \sigma}^{\dagger} L_{n m\bar{1} \nu \sigma} \nonumber\\
&\hspace{14mm} + L_{(n+1) m 1 \nu \sigma}^{\dagger} R_{n m  \bar{1} \nu \sigma}) + \text { H.c.},
\end{align}
\begin{equation}\label{eq:Hz'_integer}
\widetilde{\mathcal{H}}_{z}^{ (1)} = \sum_{\tau, \nu, \sigma} \tilde{t}_{z \nu} \: R_{n m \tau \nu \sigma}^{\dagger} L_{n m \bar{\tau} \nu \sigma} + \text { H.c.}
\end{equation}
These terms couple the remaining gapless states on the $xz$ surfaces, which become fully gapped as a result. However, two states localized to two hinges of the sample are left uncoupled: the right mover $R_{1 N_y 1 1 \bar{1}}$ at $(n,m) = (1, N_y)$ and the left mover $L_{N_z N_y \bar{1} 1 \bar{1}}$ at $(n,m) = (N_z, N_y$), see the red ovals in Fig.~\ref{fig:spectrum_l=1}. These states, which carry a single electron charge $e$, represent the chiral hinge states we are looking for.

Up to now, we have assumed that the wires are infinitely extended along the $x$ direction. However, for the sake of completeness, we should also study the properties of a finite 3D sample. We start by noting that, in the absence of interwire hopping terms along the $z$ direction, our 3D model consists of decoupled two-dimensional (2D) layers stacked along the $z$ direction. In our parameter regime of interest, $\Delta_B \gg t_y$, it has previously been shown that a single one of these layers realizes a QAH phase with a single chiral edge state propagating along the edge of the finite 2D sample~\cite{Klinovaja2015}. As such, our 3D model can be considered a stack of 2D QAH insulators stacked along the $z$ axis. Since the magnetic field, defined in Eq.~(\ref{eq:magnetic_field}), rotates oppositly in neighboring layers, these layers have opposite chiralities: in layers with $\tau = 1$ ($\tau = \bar{1}$), the edge states propagate clockwise (counterclockwise) around the finite 2D sample.

Our previous analysis for infinitely long wires showed that our system can enter a SOTI phase with a fully gapped bulk, fully gapped $xy$ and $xz$ surfaces, and two gapless chiral hinge states propagating along the $x$ direction. In the case of finite wires, one should additionally ask whether the $yz$ surfaces of the finite 3D sample are gapped or not. For this, we calculate the projections of the competing tunneling terms along the $z$ direction, $t_{z1}$ and $\tilde{t}_{z \bar{1}}$, onto the gapless QAH edge states propagating along the $y$ axis of the uncoupled 2D layers. The wave functions of these edge states can be straightforwardly found in momentum space. Let us assume for a moment that the system is periodic in the $y$ direction so that we can work with the Fourier transform characterized by the momentum $k_y$:
\begin{equation}
\Psi_{n k_y \tau \nu \sigma}(x)=\frac{1}{\sqrt{N_y}} \sum_m e^{-i m k_y a_y} \Psi_{n m \tau \nu \sigma}(x).
\end{equation}
As usual, the operator $\Psi_{n k_y \tau \nu \sigma}(x)$ can be represented in terms of slowly varying right- and left-moving fields $R_{n k_y \tau \nu \sigma}(x)$ and $L_{n k_y \tau \nu \sigma}(x)$, defined close to the Fermi points $\pm k_F$, via
\begin{equation}
\Psi_{n k_y \tau \nu \sigma} = e^{i k_F x} R_{n k_y \tau \nu \sigma} + e^{-i k_F x} L_{n k_y \tau \nu \sigma}.
\end{equation}
The Hamiltonian of the reduced system, given by $H = H_0 + H_B^{(1)} + H_y^{(1)}$, is diagonal in momentum space, such that $H = \sum_{k_y} H_{k_y}$. We start by calculating the edge state energies and wave functions at $k_y=0$ since the spectra of counterpropagating edge states are expected to intersect at this point and, hence, the gap would open here (see below). We define the Hamiltonian density $\mathcal{H}_{n \tau}$ as $H_{k_y = 0} = \sum_{n,  \tau} \int d x \, \Phi_{n \tau}^{\dagger}(x) \mathcal{H}_{n \tau} \Phi_{n \tau}(x)$, where we choose the basis $\Phi_{n \tau} = (R_{n k_y \tau 11}, L_{n k_y \tau 11} , R_{n k_y \tau 1 \bar{1}}, L_{n k_y \tau 1 \bar{1}}$, $R_{n k_y \tau \bar{1} 1}, L_{n k_y \tau \bar{1} 1}, R_{n k_y \tau \bar{1} \bar{1}}, L_{n k_y \tau \bar{1} \bar{1}})^{T}$ with $k_y = 0$. The Hamiltonian density becomes
\begin{equation}\label{eq:H_n_tau}
	\mathcal{H}_{n \tau} = v_F  \hat{k} \lambda_3 + \Delta_B[\tau  \sigma_1 \lambda_1 -  \nu_3 \sigma_2 \lambda_2]/2 + t_y \tau  \nu_1 \lambda_1,
\end{equation}
where the momentum operator $\hat{k} = - i \partial_x$ is determined near the Fermi points, and the Pauli matrices $\lambda_i$, $\sigma_i$, and $\nu_i$ for $i\in\{1,2,3\}$ act on right-/left-mover, spin, and sublattice space $\nu \in \{1, \bar{1} \}$, respectively. We then impose vanishing boundary conditions at the left and right ends of each wire. For example, at the left end of the wires $x = 0$, the wave function of the gapless edge states is set to zero: $\psi_{n \tau}(x = 0) = 0$, where $\psi_{n \tau}(x)$ is written in the basis $(\Psi_{n k_y \tau 11}, \Psi_{n k_y \tau 1 \bar{1}}$, $\Psi_{n k_y \tau \bar{1} 1}, \Psi_{n k_y \tau \bar{1} \bar{1}})^{T}$ with $k_y = 0$. From this procedure, we find that the eigenstates at $k_y=0$ have energy $E = 0$, and the corresponding wave functions take the form
\begin{equation}\label{eq:wave_function}
\begin{gathered}
	\psi_{n \tau }(x) = \\  
	\frac{1}{\sqrt{N}} \left[ 
	\begin{pmatrix}
		e^{ i \tau k_F x}\\ 
		-i  e^{-i \tau k_F x}\\
		i e^{-i \tau k_F x} \\
		-e^{i \tau k_F x} 
	\end{pmatrix}  e^{-\frac{x}{\xi_1}} + 
	\begin{pmatrix}
		-e^{- i \tau k_F x}\\
		i  e^{i \tau k_F x}\\
		-i e^{i \tau k_F x}\\
		e^{-i \tau k_F x}
	\end{pmatrix}e^{-\frac{x}{\xi_2}} \right],
\end{gathered}
\end{equation}
where $N = 2(\xi_1 + \xi_2)$ is a normalization factor, and $\xi_1 = v_F/t_y$, $\xi_2 = v_F/(\Delta_B- t_y)$ are localization lengths. At this point, we see that the strict requirement $\Delta_B \gg t_y$ (which is the regime where Fig.~\ref{fig:spectrum_l=1} is easiest to analyze) can be relaxed as long as the 2D bulk gap of the uncoupled QAH layers, given by $E_{\mathrm{gap}} = 2 \, \textup{min} [t_y, (\Delta_B - t_y)]$, remains open: $\Delta_B > t_y>0$. Furthermore, the fact that both right- and left-moving edge states are found to have energy $E = 0$ at $k_y=0$ confirms our initial assumption that their spectra cross at this point.

\begin{figure}[b]
	\centering   
	\includegraphics[width=0.47\textwidth]{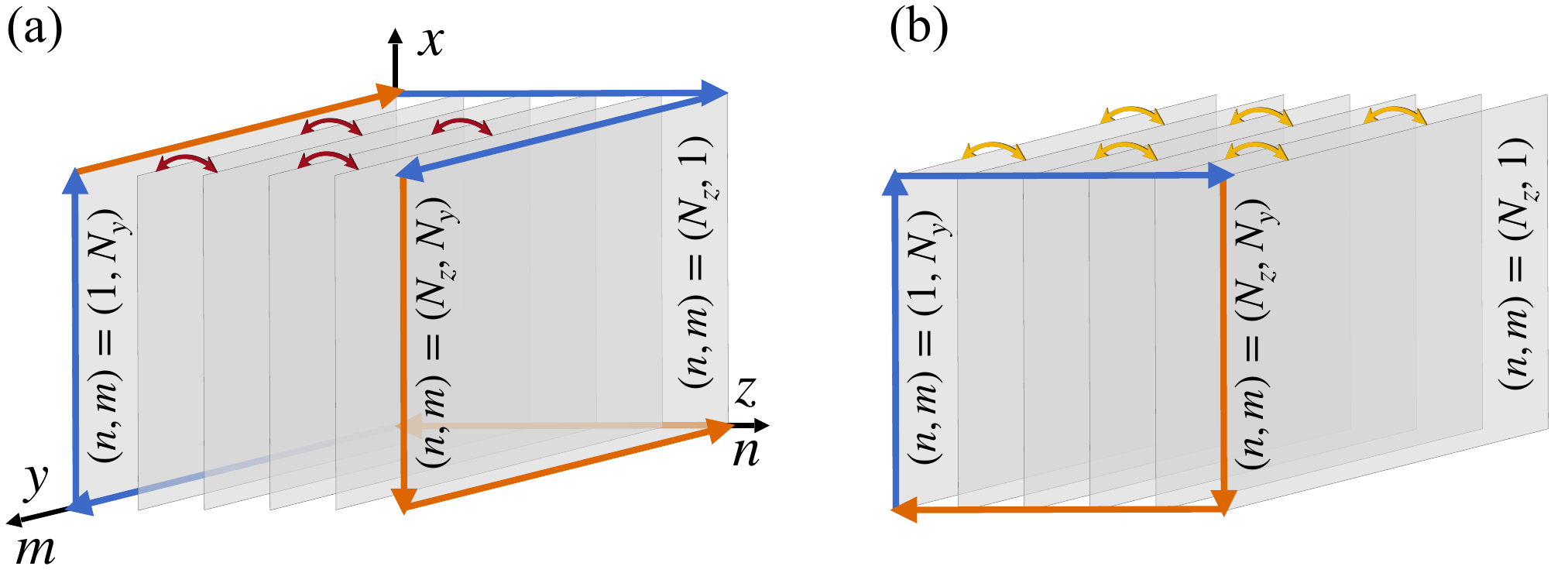}
	\caption{
		Sketch of the hinge states in the SOTI phase. Hinge states propagating to the right (left) with respect to the corresponding coordinate axis are shown in blue (orange). The hinge states follow paths that are determined by the dimerization pattern of edge states in the $yz$ plane, pictorially shown by maroon (nontrivial pattern) and yellow (trivial pattern) double arrows. The hinge states propagating along the $x$ direction are localized in the plane $m = N_y$ in agreement with Fig.~\ref{fig:spectrum_l=1}. (a) When $t_{z1} > \tilde{t}_{z \bar{1}}$, the $yz$ surfaces are gapped out nontrivially, resulting in four hinge states propagating along the $y$ axis and two hinge states propagating along the $z$ axis. (b) When $t_{z1} < \tilde{t}_{z \bar{1}}$, the $yz$ surfaces are gapped out trivially, in which case no hinge states are propagating along the $y$ direction.}
	\label{fig:dim_pat}
\end{figure}

\begin{figure*}[]
	\centering   \includegraphics[width=0.93\textwidth]{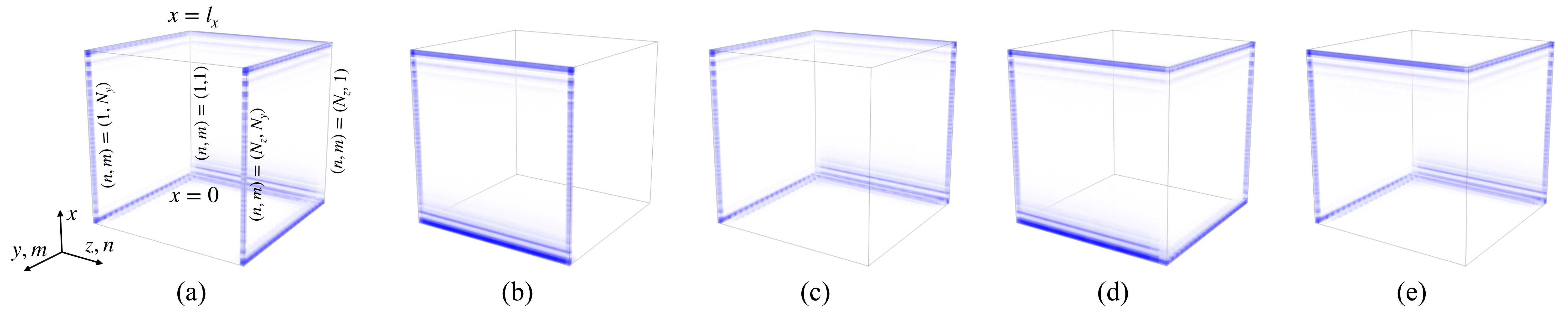}
	\caption{Probability density of the lowest-energy eigenstate calculated numerically from a discretized version of Eq.~(\ref{eq:Htotal}). In all cases, we find that hinge states (blue) propagate along closed paths of hinges consistent with the dimerization patterns and boundary terminations discussed in the main text. (a) When $t_{z1} = 0.7 \mu > \tilde{t}_{z \bar{1}} = 0.22 \mu$, the hinge state follows the path shown in Fig.~\ref{fig:dim_pat}(a). (b) When $t_{z1} = 0.22 \mu < \tilde{t}_{z \bar{1}} = 0.7 \mu$, we find ourselves in the situation illustrated in Fig.~\ref{fig:dim_pat}(b). (c) Changing the boundary termination by removing the last QAH layer leads to the relocation of one hinge state propagating along the $x$ direction from the $m = N_y$ plane to the $m=1$ plane. The hinge states propagating along the other directions adjust accordingly to create a closed hinge path in accordance with the dimerization pattern set by the parameter hierarchy $t_{z1} = 0.7 \mu > \tilde{t}_{z \bar{1}} = 0.22 \mu$. (d) The same as in (c) but with $t_{z1} = 0.22 \mu <\tilde{t}_{z \bar{1}} = 0.7 \mu$. (e) The same as in (c) but with swapped values of the $z$-tunneling amplitudes in the middle of the wires such that $t_{z1} = 0.7 \mu$, $\tilde{t}_{z \bar{1}} = 0.22 \mu$ for $x \in [0, l_x/2]$, and $t_{z1} = 0.22 \mu$, $\tilde{t}_{z \bar{1}} = 0.7 \mu$ for $x \in (l_x/2, l_x]$. The numerical parameters are $\Delta_B = 0.8 \mu$, $t_y = 0.6 \mu$, and $t_{z \bar{1}} = \tilde{t}_{z 1} = 0$. The size of the sample is $N_y \times N_z = 20 \times 20$ unit cells, and the length of the wires is $k_F l_x \approx 56$.
	}
	\label{fig:density}
\end{figure*}

We now calculate the projections of the tunneling processes along the $z$ direction [see Eqs.~(\ref{eq:Hz}) and (\ref{eq:Hz'})] onto the gapless states described by the wave functions $\psi_{n\tau}(x)$. Taking into account only the dominant terms with amplitudes $t_{z1}^{(1)}(x) = 2 t_{z1} \cos(2 k_F x)$ and $\tilde{t}_{z \bar{1}}^{(1)}(x) = 2 \tilde{t}_{z\bar{1}} \cos(2 k_F x)$, we obtain
\begin{equation}\label{eq:projection_t}
	 \left < \psi_{(n+1)1}(x) \left|t_{z1}^{(1)}(x) \frac{\mathbb{1}  + \sigma_3 \otimes  \mathbb{1}}{2} \right| \psi_{n\bar{1}}(x) \right >  = \frac{t_{z1} }{2},
\end{equation}
\begin{equation}\label{eq:projection_t_prime}
 \left < \psi_{n1}(x) \left| \tilde{t}_{z \bar{1}}^{(1)}(x) \frac{\mathbb{1} - \sigma_3 \otimes  \mathbb{1}}{2} \right| \psi_{n\bar{1}}(x) \right >  = \frac{\tilde{t}_{z\bar{1}}}{2}.
\end{equation}
We thus find that if $t_{z1} = \tilde{t}_{z\bar{1}}$, the gaps induced by these competing terms are equal such that the $yz$ surfaces remain gapless. Otherwise, if the tunneling amplitudes differ, the $yz$ surfaces are fully gapped, and only gapless hinge states remain. The path of these hinge states depends on the dimerization patterns according to which the $yz$ surfaces are gapped out: When the intercell coupling is dominant, $t_{z1} > \tilde{t}_{z\bar{1}}$, the $yz$ surfaces are gapped out in a nontrivial way, resulting in four hinge states propagating along the $y$ direction and two hinge states propagating along the $z$ direction, see Fig.~\ref{fig:dim_pat}(a). Conversely, when the intracell coupling dominates, $t_{z1} < \tilde{t}_{z\bar{1}}$, the $yz$ surfaces are gapped out in a trivial way, in which case no hinge states are propagating along the $y$ direction as shown in Fig.~\ref{fig:dim_pat}(b). In addition, we find that the previously imposed  requirement $t_y \gg t_{z1}, \tilde{t}_{z\bar{1}}$ can be relaxed as long as the bulk gap of the individual 2D layers $E_{\mathrm{gap}}$ is not closed by the tunneling processes along the $z$ axis described by Eqs.~(\ref{eq:projection_t}) and (\ref{eq:projection_t_prime}). The validity of this relaxed parameter hierarchy is further confirmed numerically.

Our analytical results can be checked numerically by exact diagonalization in the tight-binding limit; a more detailed discussion of the discretized model and some aspects of the numerical calculations are given in Appendix~\ref{App:directions}. In Figs.~\ref{fig:density}(a) and \ref{fig:density}(b), we show the probability density of the lowest-energy eigenstate obtained from a discretized version of our 3D coupled-wires model for two different values of tunneling amplitudes $t_{z1}$, $\tilde{t}_{z \bar{1}}$. In both cases, we find that the lowest-energy state is indeed tightly localized to the set of hinges highlighted in Fig.~\ref{fig:dim_pat}.

The path of the hinge states also depends on the boundary termination. To see this, we change the boundary by removing the last 2D layer lying in the $xy$ plane at $n = N_z$, $\tau = \bar{1}$. From Fig.~\ref{fig:spectrum_l=1}, it becomes clear that this modification causes one hinge state propagating along the $x$ direction to relocate from the wire with $(n,m,\tau,\nu) = (N_z, N_y, \bar{1}, 1)$ to the wire with $(N_z, 1, 1, \bar{1})$. As a consequence, the hinge states in the other directions adjust accordingly to create a closed hinge path in accordance with the corresponding dimerization pattern. see Figs.~\ref{fig:density}(c) and \ref{fig:density}(d).

Furthermore, we can obtain a more peculiar hinge path, shown in Fig.~\ref{fig:density}(e), by swapping the values of the tunneling amplitudes in the $z$ direction, $t_{z1}$ and $\tilde{t}_{z \bar{1}}$, at some point along the wires. For example, in the numerical calculations we swapped the amplitudes in the middle of the wires and used the values $t_{z1} = 0.7 \mu$, $\tilde{t}_{z \bar{1}} = 0.22 \mu$ for $x \in [0, l_x/2]$, and $t_{z1} = 0.22 \mu$, $\tilde{t}_{z \bar{1}} = 0.7 \mu$ for $x \in (l_x/2, l_x]$, where $l_x$ is the length of the wires. Although this case is rather artificial, we present it to emphasize the flexibility of our model in realizing various hinge paths by a simple adjustment of the system parameters. For completeness, we demonstrate other possible paths corresponding to different dimerization patterns and boundary terminations in systems with swapped $z$-tunneling amplitudes in Appendix~\ref{App:other}.

\begin{figure}[tb]
    \centering   \includegraphics[width=0.46\textwidth]{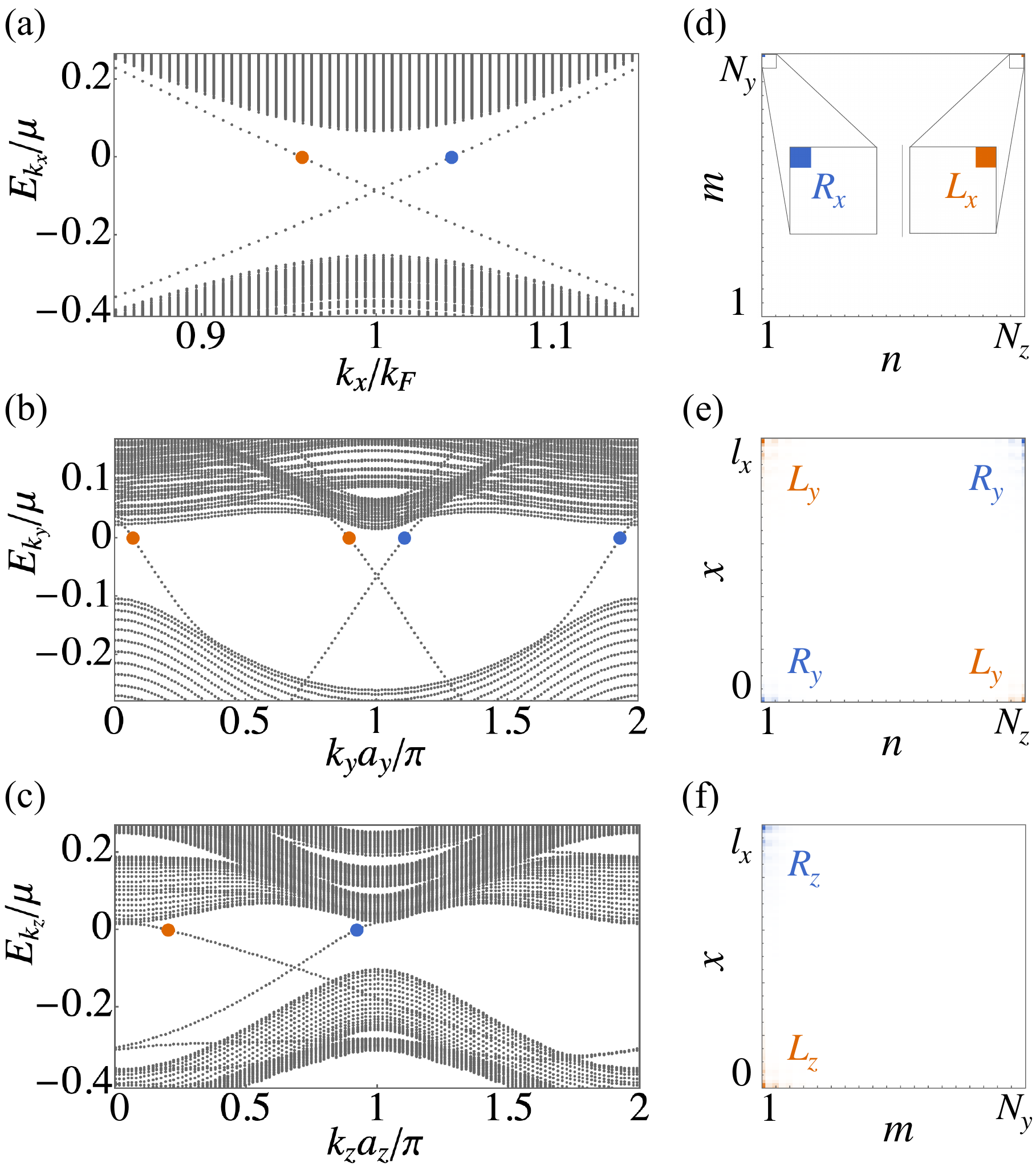}
    \caption{(a)–(c) Numerically calculated low-energy spectrum as a function of $k_x$, $k_y$, or $k_z$ for a system that is infinite along the $x$, $y$, or $z$ direction, respectively, in the parameter regime corresponding to Fig.~\ref{fig:density}(a). We see that gapless states are present in all three cases. (d)–(f) Probability density of the lowest-energy in-gap states $R_i$ ($L_i$) propagating to the right (left) along the axis $i = \{x, y, z\}$. These states are also highlighted by blue (orange) points in (a)–(c). The insets in (d) show a zoomed-in plot. Upon mapping the 2D probability densities shown here back onto a 3D sample, one recovers the hinge path shown in Fig.~\ref{fig:dim_pat}(a). The numerical parameters are the same as in Fig.~\ref{fig:density}(a).}
\label{fig:directions}
\end{figure}

We can also calculate the energy spectrum as a function of momentum $k_x$, $k_y$, or $k_z$ assuming the system is infinite along the $x$, $y$, or $z$ direction, respectively. Here, we focus on the situation shown in Fig.~\ref{fig:density}(a), but the other cases can be considered in the same way. The resulting energy spectra are shown in Figs.~\ref{fig:directions}(a)–\ref{fig:directions}(c). We find that the energy spectra shown in Figs.~\ref{fig:directions}(a) and \ref{fig:directions}(c) [Fig.~\ref{fig:directions}(b)] exhibit 2 gapless states [4 gapless states], while all other states are gapped out. In Figs.~\ref{fig:directions}(d)–\ref{fig:directions}(f), we verify that the gapless states are indeed localized to the hinges of the sample. For this, we choose the lowest-energy point on each branch and calculate the projections of the probability density of these points on the corresponding planes. Here, hinge states propagating to the right (left) with respect to the corresponding coordinate axis are shown in blue (orange). By inspection of Figs.~\ref{fig:directions}(d)–\ref{fig:directions}(f), we can again verify that right and left movers propagate along a closed path of hinges in accordance with Fig.~\ref{fig:dim_pat}(a). Note that the sizes of the energy gaps and localization lengths differ between the different panels due to the different tunneling processes that gap out the corresponding surfaces.

Finally, we checked numerically that the gapless hinge states are robust against on-site disorder of moderate strength that does not close the bulk and surface gaps. As such, the hinge states are protected by the bulk and surface gaps rather than by any special symmetries of the model; for more details, see Appendix~\ref{App:other}.

\subsection{Multiple gapless hinge states}\label{subsec:multiple}

The results obtained in the previous subsection~\ref{subsec:Single} can be generalized to the multi-state regime characterized by $l$ gapless hinge states propagating in the same direction and localized to the same hinges of the sample (here $l$ is a positive integer). Below, we consider only the case $l = 2$, and the generalization to $l>2$ is discussed in Appendix~\ref{App:multiple}.

\begin{figure}[tb]
    \centering   \includegraphics[width=0.47\textwidth]{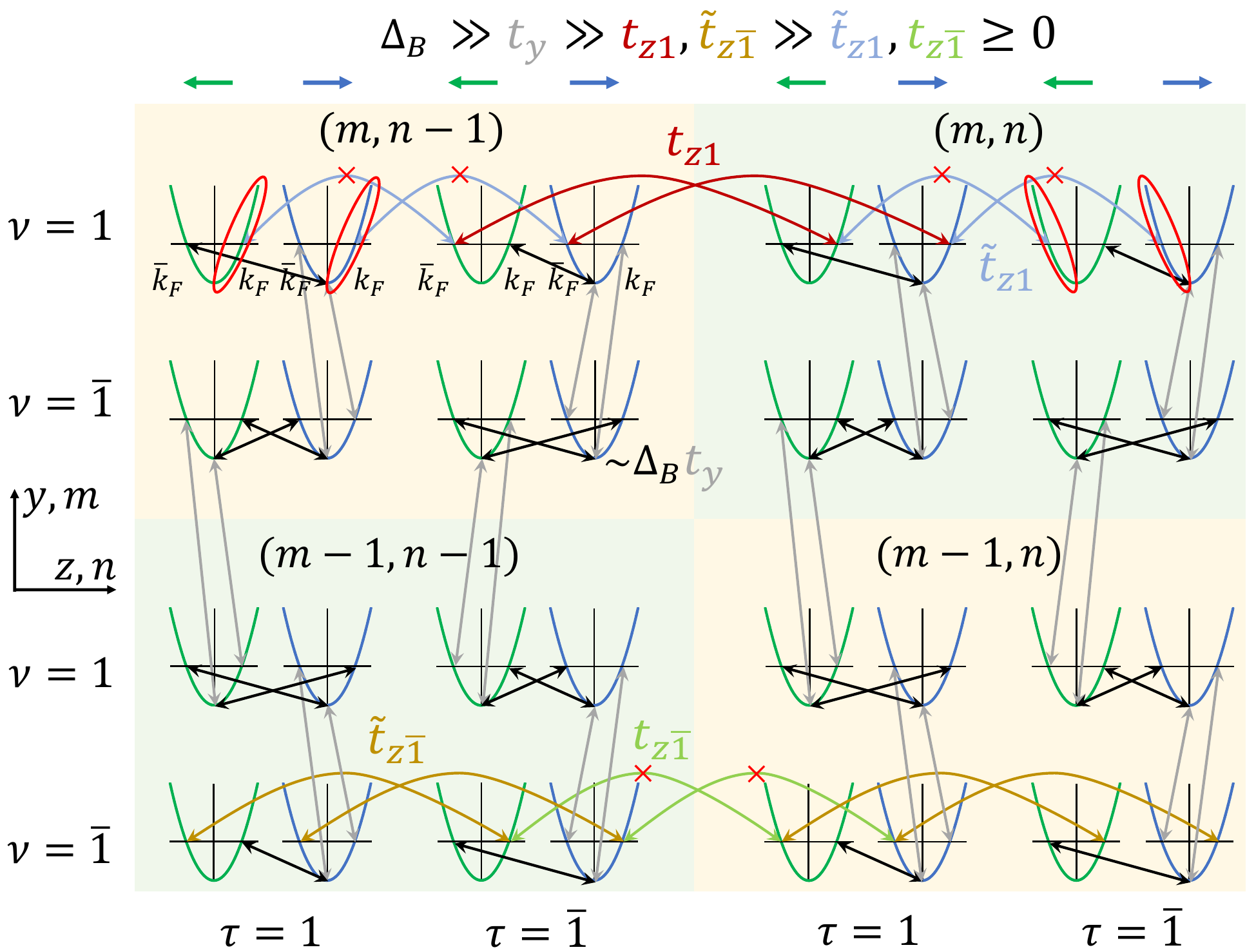}
    \caption{Sketch of four unit cells of the system with $l=2$. The notations are the same as in Fig.~\ref{fig:spectrum_l=1}. The magnetic and interwire terms in leading order of the perturbation theory open gaps in the bulk and surfaces but leave four gapless hinge states: two right movers at $(n,m,\tau, \nu) = (1, N_y, 1, 1)$, and two left movers at $(N_z, N_y, \bar{1}, 1)$. Gapless states are denoted by red ovals.}
    \label{fig:spectrum_l=2}
\end{figure}

We start by finding the hinge states propagating along the $x$ axis. For this, we again set the system to be infinite along the $x$ direction and apply a multi-step perturbation procedure. First, we take into account the Zeeman term $H_B^{(l)}$ and the tunneling along the $y$ axis term $H_y^{(l)}$, given by Eqs.~(\ref{eq:Zeeman2}) and (\ref{eq:Hy}) with $l = 2$, as these terms are assumed to be dominant in our parameter hierarchy. Without the tunneling processes along the $z$ direction, the system is again nothing but a stack of 2D QAH insulators lying in the $xy$ planes~\cite{Klinovaja2015}. The effective coupling between right and left movers in these planes is a result of two subsequent tunneling events determined in leading order of the perturbation theory with strength $\propto \Delta_B t_y/\mu $, see Fig.~\ref{fig:spectrum_l=2}. As a result, the bulk and the $xy$ surfaces of the sample become fully gapped. However, the system exhibits gapless states located on the $xz$ surfaces, namely two left movers $L_{n 1 1 \bar{1} \sigma}$ ($L_{n N_y \bar{1} 1 \sigma}$) and two right movers $R_{n N_y 1 1 \sigma}$ ($R_{n 1 \bar{1} \bar{1} \sigma}$), all with $\sigma \in \{1, \bar{1} \}$, in each 2D QAH layer at $\tau = 1$ ($\tau = \bar{1}$).

To couple the remaining gapless states but leave $2l$ chiral hinge states uncoupled, we need to slightly modify the tunneling processes along the $z$ direction. The new tunneling terms take the same form as Eqs.~(\ref{eq:Hz}) and (\ref{eq:Hz'}) with the only difference that the tunneling amplitudes do not depend on $l$. Instead they are now given by $t_{z \nu }(x) = 2 t_{z \nu} \cos \left(2 k_F x\right)$ and $\tilde{t}_{z \nu }(x) = 2 \tilde{t}_{z \nu} \cos \left(2 k_F x \right)$. These terms gap out the $xz$ surfaces but leave $l = 2$ gapless right-moving states $R_{1 N_y 1 1 \bar{1}}$, $R_{1 N_y 1 1 1}$ in the unit cell $(n, m) = (1, N_y)$, and two gapless left-moving states $L_{N_z N_y \bar{1} 1 \bar{1}}$, $L_{N_z N_y \bar{1} 1 1}$ in the unit cell $(N_z, N_y)$. These chiral hinge states are denoted by red ovals in Fig.~\ref{fig:spectrum_l=2}. Note that in the case of a finite 3D sample, the hinge states propagating along the $y$ and $z$ directions can be found in the same way as it was done for $l =1$.

\section{Fractional chiral hinge states}\label{sec:Fractional}

In this section, we show that the coupled-wires model introduced in Sec.~\ref{sec:Model} can also realize {\it fractional} SOTI phases with gapless chiral hinge states that carry only a fraction of the electronic charge $e$. For this, we set the parameter $l$ equal to $1/p$, with $p$ being a positive odd integer. In this configuration, the direct tunneling between right and left movers is suppressed due to momentum mismatch. However, we can introduce momentum-conserving tunneling terms by including backscattering terms due to strong electron-electron interactions. In the leading order that conserves momentum, the Zeeman term is modified as 
\begin{align}\label{eq:Zeeman_frac}
	\mathcal{H}_B^{(1/p)} & =  g_B \sum_{\tau} \tau 
   \left(  R_{n m \tau \tau 1}^{\dagger} L_{n m \tau  \tau \bar{1}} \left[ R_{n m \tau \tau 1}^{\dagger}  L_{n m \tau  \tau 1} \right]^q \right. \nonumber\\ 
   & \times \left[ R_{n m \tau \tau \bar{1}}^{\dagger}  L_{n m \tau  \tau \bar{1}} \right]^q  +   L_{n m \tau \bar{\tau} 1}^{\dagger} R_{n m \tau \bar{\tau} \bar{1}}  \left[ L_{n m \tau \bar{\tau} 1}^{\dagger}  \right. \nonumber \\ 
   & \left. \times R_{n m \tau \bar{\tau} 1}  \frac{}{} \right]^q \left. \left[ L_{n m \tau \bar{\tau} \bar{1}}^{\dagger}  R_{n m \tau \bar{\tau} \bar{1}}  \right]^q \right) + \text { H.c.}
\end{align}
The corresponding coupling amplitude is given by $g_B \propto \Delta_B \,g^{2 q}$, where $g$ is the strength of a single-electron backscattering process, and $q = (p-1)/2$. Similarly, we assume that tunneling processes between neighboring wires are accompanied by $q$ simultaneous backscattering processes in each wire~\cite{Kane2002,Kane2014}, such that the interwire coupling along the $y$ direction becomes
\begin{align}\label{eq:Hy_frac}
	\mathcal{H}_y^{(1/p)} & =  g_y \sum_{\tau, \sigma} \tau \left( R_{n m \tau 1 \sigma}^{\dagger} L_{n [m + (1-\sigma)/2] \tau \bar{1} \sigma} \right. \left[ R_{n m \tau 1 \sigma}^{\dagger} \right. \nonumber\\ 
	& \times \left. L_{n m \tau 1 \sigma} \frac{}{} \right]^q \left [ R_{n [m + (1-\sigma)/2] \tau \bar{1} \sigma}^{\dagger} L_{n [m + (1-\sigma)/2] \tau \bar{1} \sigma} \right ]^q   \nonumber \\ 
	&  + L_{n m \tau 1 \sigma}^{\dagger} R_{n [m + (1-\sigma)/2] \tau \bar{1} \sigma} \left [ L_{n m \tau 1 \sigma}^{\dagger} R_{n m \tau 1 \sigma} \right ]^q \nonumber \\
	& \times \left. \left[  L_{n [m + (1-\sigma)/2] \tau \bar{1} \sigma}^{\dagger}  R_{n [m + (1-\sigma)/2] \tau \bar{1} \sigma} \right]^q  \right) + \text { H.c.},
\end{align}
where $g_y \propto t_{y} \, g^{2 q}$. Hereinafter we assume that $\mathcal{H}_B^{(1/p)}$ and $\mathcal{H}_y^{(1/p)}$ are the most relevant terms in our model. This means that either (i) the bare amplitudes $g_{j}$ (with $j = B,y$) are sufficiently large, or (ii) they are quite small, such that the Hamiltonian can be studied using the standard perturbative renormalization group (RG) analysis with the scaling dimensions of $g_j$ being smaller than the scaling dimensions of all possible competing terms. As pointed out in Ref.~\cite{Kane2002}, it is usually possible to construct such interactions. In addition, we assume that $g_y$ is weak compared to $g_B$.

In the same way, we combine Eqs.~(\ref{eq:Hz}) and (\ref{eq:Hz'}) with backscattering processes such that momentum-conserving interwire coupling terms along the $z$ direction are obtained. In leading order in the interactions, these terms take the form
\begin{align}\label{eq:Hz_frac}
	\mathcal{H}^{(1/p)}_z & = \sum_{\nu, \sigma} g_{z \nu} \left( R_{(n+1) m 1 \nu \sigma}^{\dagger} L_{n m \bar{1} \nu \sigma} \right. 
	\nonumber\\ 
	& \times  \left[ R_{(n+1) m 1 \nu \sigma}^{\dagger} L_{(n+1) m 1 \nu \sigma} \frac{}{} \right]^q  \left[ R_{n m\bar{1} \nu \sigma}^{\dagger} L_{n m\bar{1} \nu \sigma} \right]^q  
	\nonumber\\ 
	&  + L_{(n+1) m 1 \nu \sigma}^{\dagger} R_{n m  \bar{1} \nu \sigma} \left[ L_{(n+1) m 1 \nu \sigma}^{\dagger} R_{(n+1) m 1 \nu \sigma} \right]^q  
	\nonumber\\ 
	&  \times \left. \left[ L_{n m  \bar{1} \nu \sigma}^{\dagger} R_{n m  \bar{1} \nu \sigma} \right]^q  \right) + \text { H.c.},
\end{align}
\begin{align}\label{eq:Hz'_frac}
	\widetilde{\mathcal{H}}_{z}^{(1/p)} & = \sum_{\tau, \nu, \sigma} \tilde{g}_{z \nu} \left( R_{n m \tau \nu \sigma}^{\dagger} L_{n m \bar{\tau} \nu \sigma} \left[ R_{n m \tau \nu \sigma}^{\dagger} L_{n m \tau \nu \sigma} \right]^q \right.
	\nonumber\\ 
	& \times \left. \left[R_{n m \bar{\tau} \nu \sigma}^{\dagger} L_{n m \bar{\tau} \nu \sigma} \right]^q \right) + \text { H.c.}
\end{align}
The modified coupling amplitudes are determined by $g_{z \nu} \propto t_{z \nu} \, g^{2 q}$, $\tilde{g}_{z \nu} \propto \tilde{t}_{z \nu}\, g^{2 q}$, and the terms $\mathcal{H}^{(1/p)}_z$ and $\widetilde{\mathcal{H}}_{z}^{ (1/p)}$ are assumed to be relevant in the RG sense. Similarly to the integer case, we focus on the parameter hierarchy $g_y \gg g_{z 1}, \tilde{g}_{z \bar{1}} \gg g_{z \bar{1}}, \tilde{g}_{z 1} \geq 0$.

To simplify the above expressions, we can introduce new right- and left-moving field of the form~\cite{Sagi2014}
\begin{align}\label{eq:tilde_R_L}
	\widetilde{R}_{n m \tau \nu \sigma} & = \left[ R_{n m \tau \nu \sigma} \right]^{q+1} \left[ L_{n m \tau \nu \sigma}^{\dagger} \right]^{q},
	\nonumber \\ 
	\nonumber \\ 
	\widetilde{L}_{n m \tau \nu \sigma} & = [L_{n m \tau \nu \sigma}]^{q+1} [R_{n m \tau \nu \sigma}^{\dagger}]^{q}.
\end{align}
In terms of these new fields, the Hamiltonian terms defined in Eqs.~(\ref{eq:Zeeman_frac})--(\ref{eq:Hz'_frac}) take the same form as their `integer' counterparts given by Eqs.~(\ref{eq:Zeeman_integer})--(\ref{eq:Hz'_integer}) for $l = 1$. Thus, by analogy with the integer case, we can conclude the following in leading order of the perturbation theory: First, the states $\widetilde{R}_{1 N_y 1 1 \bar{1}}$ and $\widetilde{L}_{N_z N_y \bar{1} 1 \bar{1}}$ do not enter the sum $H_B^{(1/p)} + H_y^{(1/p)} + H^{(1/p)}_z + \widetilde{H}_{z}^{(1/p)}$, which makes them potential candidates for the fractional gapless hinge states we are looking for. Second, all other states $\widetilde{R}_{n m \tau \nu \sigma}$ and $\widetilde{L}_{n m \tau \nu \sigma}$ enter the sum exactly once, ensuring that all terms in the sum commute with each other and potentially induce gaps in the bulk and surfaces. To make this more explicit, we can apply the standard bosonization procedure (see, e.g., Ref.~\cite{Giamarchi}) to our system. We thus express the right and left movers in terms of the bosonic fields $\varphi_{r n m \tau \nu \sigma }(x)$ as 
\begin{align}
R_{n m \tau \nu \sigma}(x) & \propto e^{i \varphi_{1  n m \tau \nu \sigma}(x)}, \nonumber \\ 
L_{n m \tau \nu \sigma}(x) & \propto e^{i \varphi_{\bar{1}
  n m \tau \nu \sigma}(x)}.
\end{align}
These new fields obey the commutation relations
\begin{align}
& {\left[\varphi_{r n m \tau \nu \sigma}(x), \varphi_{r^{\prime} n^{\prime} m^{\prime} \tau^{\prime} \nu^{\prime} \sigma^{\prime} }\left(x^{\prime}\right)\right]} \nonumber \\
& \quad=i r \pi \delta_{r r^{\prime}} \delta_{n n^{\prime}} \delta_{m m^{\prime}}  \delta_{\tau \tau^{\prime}} \delta_{\nu \nu^{\prime}} \delta_{\sigma \sigma^{\prime}} \operatorname{sgn}\left(x-x^{\prime}\right),
\end{align}
with $r \in \{1, \bar{1} \}$. These relations, together with an appropriate choice of Klein factors, ensure the anticommutation relations of the fermionic fields~\cite{Giamarchi}. However, for our purposes, the Klein factors can be safely ignored~\cite{Kane2014}.

In bosonized language, the new right- and left-moving fields defined in Eq.~(\ref{eq:tilde_R_L}) take the form $\widetilde{R}_{n m \tau \nu \sigma}(x)  \propto e^{i \widetilde{\varphi}_{1  n m \tau \nu \sigma}(x)}$ and $\widetilde{L}_{n m \tau \nu \sigma}(x)  \propto e^{i \widetilde{\varphi}_{\bar{1} n m \tau \nu \sigma}(x)}$, where we have introduced new bosonic fields
\begin{equation}
	\widetilde{\varphi}_{r n m \tau \nu \sigma} = (q + 1) \: \varphi_{r n m \tau \nu \sigma} - q \: \varphi_{\bar{r} n m \tau \nu \sigma},
\end{equation}
satisfying nontrivial commutation relations
\begin{align}\label{eq:eta}
& \left[ \widetilde{\varphi}_{r n m \tau \nu \sigma}(x),  \widetilde{\varphi}_{r^{\prime} n^{\prime} m^{\prime} \tau^{\prime} \nu^{\prime} \sigma^{\prime} } \left(x^{\prime}\right) \right] 
\nonumber \\
& \quad = i p r \pi \delta_{r r^{\prime}} \delta_{n n^{\prime}} \delta_{m m^{\prime}} \delta_{\tau \tau^{\prime}} \delta_{\nu \nu^{\prime}} \delta_{\sigma \sigma^{\prime}} \operatorname{sgn}\left(x-x^{\prime}\right).
\end{align}
When expressed using these new bosonic fields, the Hamiltonian terms given in Eqs.~(\ref{eq:Zeeman_frac})--(\ref{eq:Hz'_frac}) take the form
\begin{align}\label{eq:Zeeman_bose}
	\mathcal{H}_B^{(1/p)} 
	& \propto g_B \sum_{\tau} \tau 
	\left[ \right. \cos(\widetilde{\varphi}_{1 n m \tau \tau 1} - \widetilde{\varphi}_{\bar{1} n m \tau  \tau \bar{1}}) 
	\nonumber \\
	& + \cos( \widetilde{\varphi}_{\bar{1} n m \tau \bar{\tau} 1} -  \widetilde{\varphi}_{1 n m \tau \bar{\tau} \bar{1}}) \left.\right],
\end{align}
\begin{align}
	\mathcal{H}_y^{(1/p)} 
	& \propto g_y \sum_{\tau, \sigma} \tau \left[ \cos(\widetilde{\varphi}_{1 n m \tau 1 \sigma} - \widetilde{\varphi}_{\bar{1} n [m + (1-\sigma)/2] \tau \bar{1} \sigma}) \right. 
	\nonumber \\
	& + \cos(\widetilde{\varphi}_{\bar{1} n m \tau 1 \sigma} -  \widetilde{\varphi}_{1 n [m + (1-\sigma)/2] \tau \bar{1} \sigma})\left. \right],
\end{align}
\begin{align}
	\mathcal{H}^{(1/p)}_z
	& \propto \sum_{\nu, \sigma} g_{z \nu} \left[ \right. \cos( \widetilde{\varphi}_{1 (n+1) m 1 \nu \sigma} -  \widetilde{\varphi}_{\bar{1} n m\bar{1} \nu \sigma})  
	\nonumber \\
	& + \cos(\widetilde{\varphi}_{\bar{1} (n+1) m 1 \nu \sigma} - \widetilde{\varphi}_{1 n m  \bar{1} \nu \sigma}) \left. \right], 
\end{align}
\begin{equation}\label{eq:Hz'_bose}
	\widetilde{\mathcal{H}}_{z}^{(1/p)} \propto \sum_{\tau, \nu, \sigma} \tilde{g}_{z \nu} \cos(\widetilde{\varphi}_{1 n m \tau \nu \sigma} -  \widetilde{\varphi}_{\bar{1} n m \bar{\tau} \nu \sigma}).
\end{equation}
The terms given in Eqs.~(\ref{eq:Zeeman_bose})-(\ref{eq:Hz'_bose}) pin the arguments of the cosines to constant values in order to minimize the total energy of the system~\cite{Laubscher2023,Kane2002,Kane2014,Klinovaja2014b,Sagi2015b,Klinovaja2015,Klinovaja2014,Sagi2014,Santos2015,Sagi2015,Meng2015}. In the parameter regime we are considering, all fields in the bulk and on the surfaces of the sample are pinned in a pairwise fashion, such that the bulk and surfaces are fully gapped. However, two chiral states $\widetilde{\varphi}_{1 1 N_y 1 1 \bar{1}}$ and $\widetilde{\varphi}_{\bar{1} N_z N_y \bar{1} 1 \bar{1}}$, propagating along the $x$ direction and localized at the hinges of the sample, do not appear in the sums and thus remain gapless. Following Refs.~\cite{Kane2002,Kane2014}, one can show that these states carry a fractional charge $e/p$.

Finally, we discuss the fractional gapless hinge states propagating along the $y$ and $z$ directions of a finite 3D sample. As was mentioned above, in terms of the fields $\widetilde{R}_{n m \tau \nu \sigma}$ and $\widetilde{L}_{n m \tau \nu \sigma}$ defined in Eq.~(\mbox{\ref{eq:tilde_R_L}}), the fractional Hamiltonian takes the same form as the integer Hamiltonian with $l=1$. As a result, the fractional gapless states are expected to form hinge paths in the same manner as their integer counterparts~\cite{Laubscher2023,Kane2002,Kane2014,Klinovaja2014b,Sagi2015b,Klinovaja2015,Klinovaja2014,Sagi2014,Santos2015,Sagi2015,Meng2015}. Again, the hinge paths can be controlled by the boundary termination of the sample and the hierarchy of the coupling amplitudes (see Figs.~\ref{fig:dim_pat} and~\ref{fig:density}).

\section{CONCLUSIONS AND OUTLOOK}\label{sec:conclusion}

We have constructed a model of a chiral 3D SOTI from an array of weakly coupled nanowires. The specific choice of helical magnetic fields and spatially modulated interwire couplings allows the model to host (single or multiple) integer or fractional gapless chiral hinge states, while the bulk and surfaces remain fully gapped. The fractional regime emerges in the presence of strong electron-electron interactions, which have been effectively treated using a bosonized language. In this regime, the hinge states carry a fraction of the elementary electron charge $e/p$ (with $p$ being an odd positive integer), and quasiparticle excitations are predicted to obey nontrivial Abelian braiding statistics~\cite{Kane2002,Kane2014}. Furthermore, the path of the gapless hinge states can be controlled by adjusting the interwire coupling amplitudes and boundary terminations.

We have proposed two potential realizations of the model: one of them is based on rotating magnetic fields, and the other one on the interplay between SOI and uniform magnetic fields. In both cases, the total magnetization is equal to zero, such that the SOTI phases constructed in this work effectively correspond to integer and fractional QAH phases. Although our model is primarily of theoretical interest and serves to expand the set of analytically tractable toy models for strongly interacting phases, some of its aspects can, in principle, be implemented in experiments. In this context, recent progress toward the experimental realization of the fractional QAH effect in layered systems~\cite{Cai2023, Zeng2023, Park2023, Xu2023, Lu2024} is encouraging. We expect that the fractionally charged hinge states of our SOTI can be detected via similar methods as the fractionally charged edge states in these works, such as, e.g., magneto-optically observed Landau-fan diagrams~\cite{Cai2023}, a combination of local electronic compressibility and magneto-optical measurements~\cite{Zeng2023}, or the observation of fractional plateaus of the quantized Hall resistance~\cite{Park2023, Xu2023, Lu2024}.

Finally, we note that our 3D model of coupled wires could potentially be modified to enter other second-order QAH phases at filling factors $l = q/p$, where $q$ and $p$ are positive integers. This modification would allow the system to host not only Abelian but also non-Abelian quasiparticle excitations.

\acknowledgments
This work was supported by the Swiss National Science Foundation and NCCR SPIN (Grant No. 51NF40-180604). This project received funding from the European Union’s Horizon 2020 research and innovation program (ERC Starting Grant, Grant Agreement No. 757725). K. L. acknowledges support by the Laboratory for Physical Sciences through the Condensed Matter Theory Center.

\section*{Data availability}
The data that support the findings of this article are available in the following Zenodo repository: \href{https://doi.org/10.5281/zenodo.15857796}{10.5281/zenodo.15857796}.

\appendix

\section{Hinge states in the model with SOI}\label{App:SOI}

In this Appendix, we show that the SOI-based model, described in subsection~\ref{subsec:SOI} of the main text, is equivalent to the model based on helical magnetic fields and, hence, can also host gapless chiral hinge states. In the model with SOI, the chemical potential is tuned to $\mu^{(l)} = E_{so} (l^2 - 1)$ and the Fermi momenta are given by $k_{F\pm}^{(l)} = k_{so}(1 \pm l)$. In the usual manner, we rewrite the Hamiltonian in a basis of slowly varying right and left movers inside each wire:
\begin{equation}
	\Psi_{n m \tau \nu (\tau \bar{\nu})} =e^{i k_{F+}^{(l)} x} R_{n m \tau \nu (\tau \bar{\nu})}+ e^{i k_{F-}^{(l)} x} L_{n m \tau \nu (\tau \bar{\nu})},
\end{equation}
\begin{equation}
	\Psi_{n m \tau \nu (\tau \nu)} =  e^{-i k_{F-}^{(l)} x} R_{n m \tau \nu (\tau \nu)}+ e^{-i k_{F+}^{(l)} x} L_{n m \tau \nu (\tau \nu)}.
\end{equation}  
As a result, we find the Hamiltonian density
\begin{align}
	\mathcal{H}_0 + \mathcal{H}_{SOI} &= -i v_F^{(l)} \sum_{\tau, \nu, \sigma}  (R_{n m \tau \nu \sigma}^{\dagger}  \partial_x R_{n m \tau \nu \sigma} \nonumber\\
	&\hspace{20mm}-L_{n m \tau \nu \sigma}^{\dagger} \partial_x L_{n m \tau \nu \sigma} ),
\end{align}
where $v_F^{(l)} = \alpha l$. This expression coincides with the Hamiltonian density $\mathcal{H}_0$ of the main model given by Eq.~(\ref{eq:H0}). It is then straightforward to show that for  $l = 1$, the magnetic and tunneling terms written in the basis of right and left movers are the same as those in Fig.~\ref{fig:spectrum_l=1} in leading order of the perturbation theory. Hence, the alternative model has a fully gapped bulk as well as fully gapped $xy$ and $xz$ surfaces, but hosts two gapless hinge states $R_{1 N_y 1 1 \bar{1}}$ and $L_{N_z N_y \bar{1} 1 \bar{1}}$. 

Similarly, it can easily be shown that the results for the multi-state regime from subsection~\ref{subsec:multiple} are also valid for the SOI-based model. However, in contrast to the cases of $l=1$ and $l = 1/p$, the tunneling amplitudes in the $z$ direction now depend on $l$ as $t_{z \nu }^{(l)}(x) = 2 t_{z \nu } [\cos(2 k_{F+}^{(l)} x) + \cos(2 k_{F-}^{(l)}x)]$ and $\tilde{t}_{z \nu }^{(l)}(x) = 2 \tilde{t}_{z \nu } [\cos(2 k_{F+}^{(l)} x) + \cos(2 k_{F-}^{(l)}x)]$. In this case, they couple both exterior modes with Fermi momenta $\pm k_{F+}^{(l)}$ and interior modes with $\pm k_{F-}^{(l)}$, but leave $2 l$ gapless hinge states.

\section{Tight-binding model}\label{App:directions}

In this Appendix, we discuss the discretized version of our model and some details of the numerical calculations. In the tight-binding limit, we assume that each wire is composed of a finite number of sites located at $x_j = a_x j$, where $a_x$ is the distance between two neighboring sites and $j$ enumerates the sites. We then introduce the creation and annihilation operators of an electron on each site $j$, $(\Psi_{n m \tau \nu \sigma}^{j})^\dagger$ and $\Psi_{n m \tau \nu \sigma}^{j}$. The kinetic Hamiltonian term, defined in Eq.~(\ref{eq:kinetic}), can be straightforwardly rewritten in the basis of these new operators using a finite difference method. This yields 
\begin{align}
H_0 = & \sum_{n, m} \sum_{\tau, \nu, \sigma} \sum_j \big\{(\Psi_{n m \tau \nu \sigma}^{j})^\dagger \left[ 2 t_x - \mu \right] \Psi_{n m \tau \nu \sigma}^j \nonumber\\ & - \left[ t_x (\Psi_{n m \tau \nu \sigma}^{j})^\dagger  \Psi_{n m \tau \nu \sigma}^{j-1} + \text{H.c.} \right]\big\},
\end{align}	
where $t_x = 1/(2 m_0 a_x^2)$ is a hopping amplitude between adjacent sites. The magnetic and interwire tunneling terms, given by Eqs.~(\ref{eq:Zeeman2})--(\ref{eq:Hz'}), can be rewritten as
\begin{align}
    H_{B}^{(l)} = & \Delta_B  \sum_{n, m} \sum_{\tau, \nu}  \tau \sum_j  (\Psi_{n m  \tau \nu 1}^{j})^\dagger    \nonumber\\ &\quad\times  \Psi^{j}_{n m \tau  \nu \bar{1}} \exp \left[ i (\tau \nu)  \: 2 k_{F} \frac{x_j}{l} \right] + \text{H.c.},
\end{align}	
\begin{align}
    H_y^{(l)} = \sum_{n,m} \sum_{\tau, \sigma} \tau &  \sum_j 
    t^{(l)}_y(x_j) (\Psi_{n m \tau 1 \sigma}^{j})^\dagger  \nonumber\\ &\quad\times \Psi_{n [m + (1-\sigma)/2]  \tau \bar{1} \sigma}^j+ \text{ H.c.},
\end{align}	
\begin{equation}
H_z^{(l)} = \sum_{n, m} \sum_{\nu, \sigma} \sum_j  t_{z\nu}^{(l)}(x_j) (\Psi_{(n+1) m 1 \nu \sigma}^{j})^\dagger \Psi_{n m \bar{1} \nu \sigma}^j+\text { H.c.},
\end{equation}
\begin{equation}
\widetilde{H}_{z}^{(l)} = \sum_{n, m} \sum_{\nu, \sigma} \sum_j \tilde{t}_{z \nu }^{(l) }(x_j) (\Psi_{n m 1 \nu \sigma}^{j})^\dagger \Psi_{n m \bar{1} \nu \sigma}^j+\text { H.c.}
\end{equation}
Taken together, these Hamiltonian terms represent a discretized version of the continuous model of Eq.~(\ref{eq:Htotal}) that we use in our numerical calculations.

Next, we discuss the details of how to calculate the spectrum and probability density of the hinge states propagating along the $x$ direction, presented in Figs.~\ref{fig:directions}(a) and \ref{fig:directions}(d) of the main text. First, we assume that the system is periodic in the $x$ direction with period $\tilde{a}_x = n_x a_x = \pi/k_F$, where $n_x$ is an integer. We can thus work with the Fourier transform characterized by the momentum $k_x$:
\begin{equation}\label{eq:psi_kx}
\Psi_{n m \tau \nu \sigma}^{k_x}=\frac{1}{\sqrt{N_x}} \sum_{\eta=1}^{N_x} e^{-i \eta k_x \tilde{a}_x} \Psi_{n m \tau \nu \sigma}^\eta,
\end{equation}
where $N_x$ is the number of unit cells (one unit cell contains $n_x$ sites), $k_x$ takes values within the Brillouin zone $-\pi/\tilde{a}_x < k_x < \pi/\tilde{a}_x$, and $\Psi_{n m \tau \nu \sigma}^\eta$ consists of the operators on each site within the unit cell: 
\begin{equation}
	\Psi_{n m \tau \nu \sigma}^\eta = \left( \Psi_{n m \tau \nu \sigma}^{j = 1 + \delta_{\eta}},
	\Psi_{n m \tau \nu \sigma}^{j = 2+\delta_{\eta}}, ..., 
	\Psi_{n m \tau \nu \sigma}^{j = n_x + \delta_{\eta}} \right)^T,
\end{equation}
where $\delta_\eta = n_x (\eta - 1)$. Next, we can rewrite the Hamiltonian in the basis of the Fourier-transformed fields $\Psi_{n m \tau \nu \sigma}^{k_x}$ and find its eigenenergies $E_{k_x}$ and eigenfunctions $\phi_{k_x}(y, z)$ in dependence on $k_x$. This allows us to calculate the probability density $|\phi_{k_x}(y,z)|^2$ of the right- and left-moving hinge state at energy $E_{k_x} = 0$ for the corresponding values of $k_x$.

Similarly, we can examine the hinge states
propagating along the $y$ and $z$ axes. For this, we assume that the system is periodic in the $y$ or $z$ direction, respectively, with a period $a_y$ or $a_z$. We then pass to momentum space characterized by the momentum $k_y$ or $k_z$, respectively, via the Fourier transform

\begin{equation}
\Psi_{n k_y \tau \nu \sigma}^j = \frac{1}{\sqrt{N_y}} \sum_m e^{-i m k_y a_y} \Psi_{n m \tau \nu \sigma}^j, 
\end{equation}

\begin{equation}
\Psi_{k_z m \tau \nu \sigma}^j = \frac{1}{\sqrt{N_z}} \sum_n e^{-i n k_z a_z} \Psi_{n m \tau \nu \sigma}^j.
\end{equation}

\section{Hinge paths in systems with swapped values of the $z$-tunneling amplitudes. Influence of disorder}\label{App:other}

\begin{figure*}[]
	\centering   \includegraphics[width=0.99\textwidth]{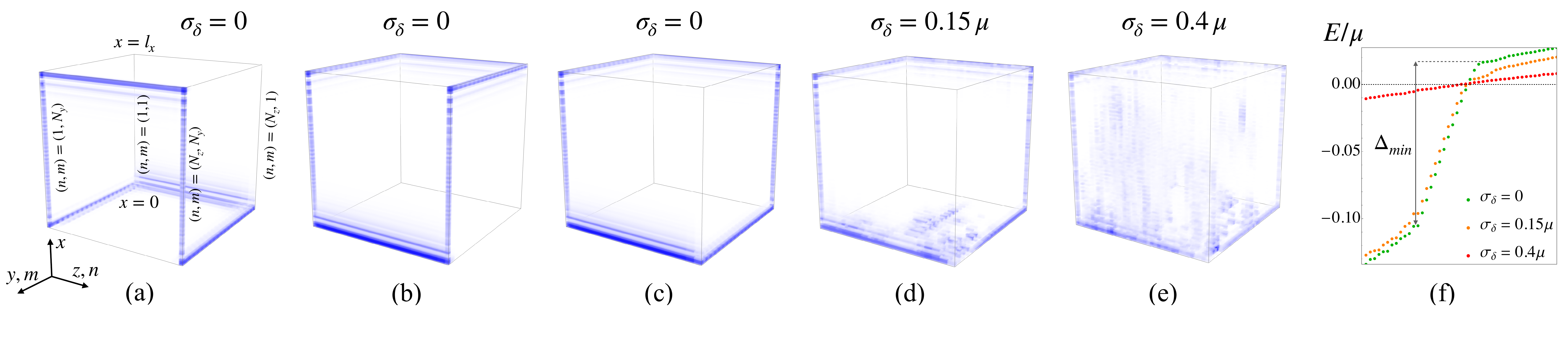}
	\caption{Probability density of the lowest-energy eigenstate [panels (a)--(e)] and low-energy spectrum [panel (f)] calculated numerically from a discretized version of Eq.~(\ref{eq:Htotal}). Panels (a)--(c) have the same dimerization patterns and boundary terminations as the ones in Figs.~\ref{fig:density}(a), (b), and (d) of the main text, respectively. The key difference from Figs.~\ref{fig:density}(a), (b), and (d) is that the values of the $z$-tunneling amplitudes are swapped in the middle of the wires such that in (a) we have $t_{z1} = 0.7 \mu$, $\tilde{t}_{z \bar{1}} = 0.22 \mu$ for $x \in [0, l_x/2]$, and $t_{z1} = 0.22 \mu$, $\tilde{t}_{z \bar{1}} = 0.7 \mu$ for $x \in (l_x/2, l_x]$; in (b), (c) we have $t_{z1} = 0.22 \mu$, $\tilde{t}_{z \bar{1}} = 0.7 \mu$ for $x \in [0, l_x/2]$, and $t_{z1} = 0.7 \mu$, $\tilde{t}_{z \bar{1}} = 0.22 \mu$ for $x \in (l_x/2, l_x]$. Panels (d) and (e) show the same system as (c) but with additional on-site disorder of strength $\sigma_\delta$. We find that the hinge states are robust against disorder of moderate strength $\sigma_\delta = 0.15 \, \mu$; see panel (d). However, an increased disorder strength of $\sigma_\delta = 0.4 \, \mu$ leads to the closing of the gap [panel (f)] and delocalization of the hinge states [panel (e)]. The numerical parameters are the same as in Fig.~\ref{fig:density}.}
	\label{fig:paths}
\end{figure*}

For the sake of completeness, in Figs.~\ref{fig:paths}(a)-~\ref{fig:paths}(c), we present possible paths of hinge states in systems where the values of the tunneling amplitudes along the $z$ direction, $t_{z1}$ and $\tilde{t}_{z \bar{1}}$, are swapped in the middle of the wires. These examples serve to demonstrate the flexibility of our model to generate different hinge paths. The different panels correspond to different dimerization patterns and boundary terminations.

Next, we investigate numerically how on-site disorder in the chemical potential affects the hinge states shown in Fig.~\ref{fig:paths}(c).  For this, we add disorder $\delta$ at each site of the system, where the random variable $\delta$ follows a Gaussian distribution with zero mean $\left< \delta \right> = 0$ and standard deviation $\sigma_\delta$. We find that the gapless hinge states are robust against disorder of moderate strength as long as the bulk and surface gaps remain open. As shown in Fig.~\ref{fig:paths}(d), the hinge states survive for a disorder strength of $\sigma_\delta=0.15 \, \mu$, even though the value of $\sigma_\delta$ exceeds the minimum gap $\Delta_{\min}\approx0.12\,\mu$ found from Fig.~\ref{fig:paths}(f). However, increasing the disorder strength further closes the gap and, as a result, destroys the hinge states; see Figs.~\ref{fig:paths}(e) and~\ref{fig:paths}(f) for $\sigma_\delta = 0.4 \, \mu$. Note that the other cases shown in Fig.~\ref{fig:density}, Fig.~\ref{fig:paths}(a), and Fig.~\ref{fig:paths}(b) can be treated in the same way. Generally, we find that the hinge states are stable against disorder and do not require any spatial symmetries to be present.

\section{Generalization to multiple edge modes ($l>2$)}\label{App:multiple}

In Sec.~\ref{sec:Integer} of the main text, we have studied the Hamiltonian introduced in Eq.~(\ref{eq:Htotal}) for $l=1$ and $l=2$. In both cases, we found that the model hosts $l$ gapless hinge states propagating in the same direction and localized to the same hinges of the sample. In this Appendix, we discuss how to modify our model for other positive integer values of the parameter $l>2$.

We focus on gapless hinge states propagating along the $x$ direction in a system that is assumed to be infinite along the $ x $ direction. We start by considering the Zeeman term $H_B^{(l)}$ and the tunneling term along the $y$ direction $H_y^{(l)}$ as these are assumed to be dominant in our parameter hierarchy. These terms are given by Eqs.~(\ref{eq:Zeeman2}) and (\ref{eq:Hy}) of the main text, where $l$ is now an arbitrary positive integer larger than one. In this case, the effective coupling between right and left movers results from $l$ sequential tunneling events with strengths determined to leading order in perturbation theory as  $\propto t_y^{l_1} \Delta_B^{l - l_1}/\mu^{l-1}$. Here, $l_1$ assumes values $\{ 1, 2,..., l-1\}$. For example, from Fig.~\ref{fig:spectrum_l=2} of the main text shown for $l = 2$, we can read off that $l_1 = 1$. Focusing on the multi-state regime with $l>2$ (for simplicity, on the case $l = 3$), we find coupling strengths proportional to $ t_y \Delta_B^2/ \mu^2$ with $l_1 = 1$, and proportional to $t_y^2 \Delta_B/ \mu^2$ with $l_1 = 2$.

\begin{figure}[tb]
    \centering   \includegraphics[width=0.47\textwidth]{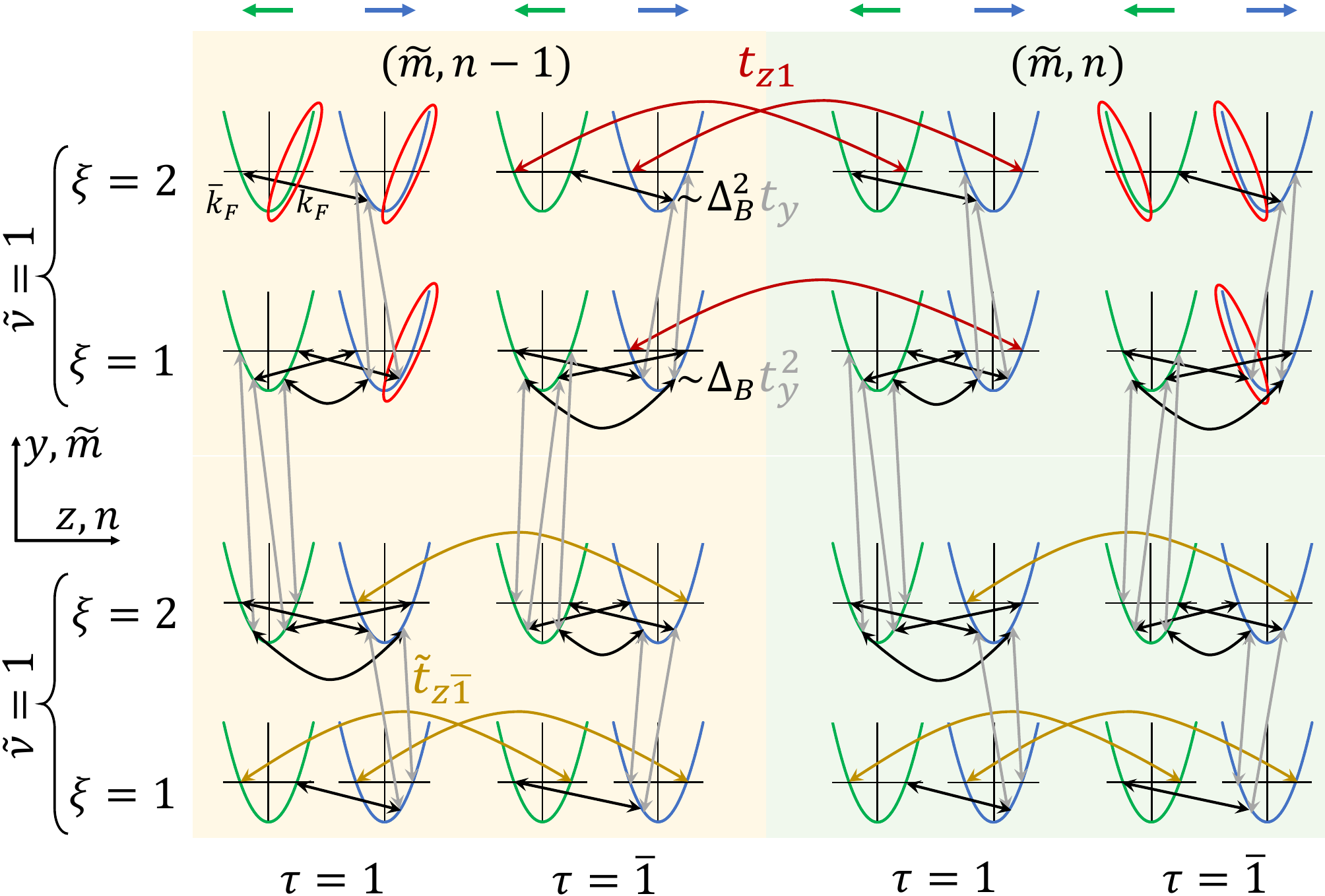}
    \caption{Sketch of two unit cells of the coupled-wires model with $l=3$. The notations are the same as in Fig.~\ref{fig:spectrum_l=1}. The magnetic and interwire terms in leading order of the perturbation theory open gaps in the bulk and surfaces but leave $2 l$ gapless hinge states (denoted by red ovals): three right movers at $(n, \tilde{m},\tau, \tilde{\nu}) = (1, N_y, 1, 1)$, and three left movers at $(N_z, N_y, \bar{1}, 1)$.}
    \label{fig:spectrum_l=3}
\end{figure}

To progress further, we change the structure of the unit cells in the system. We combine $\eta_{l}/4$ old unit cells that are adjacent to each other in the $y$ direction into one new unit cell, such that the new unit cell consists of $\eta_l$ wires, where 
\begin{equation}
    \eta_l = 
 \left\{\begin{matrix}
2l + 2, & \text{for odd } l, \\
2 l,  & \text{for even } l,\\
\end{matrix}\right.
\end{equation}
see Fig.~\ref{fig:spectrum_l=3} for a schematic illustration in the case $l=3$. As before, the index $n$ denotes the position of a unit cell along the $z$ direction, and $\tau \in \{1, \bar{1} \}$ is used to label wires within a unit cell, representing their positions relative to the $z$ axis as left or right. The position of a unit cell along the $y$ direction is now indicated by the index $\tilde{m}$. Furthermore, the position of a wire within a unit cell in the $y$ direction is denoted by two indices $(\tilde{\nu}, \xi)$. Here, $\tilde{\nu} \in \{1, \bar{1} \}$ represents not a single layer in the $xz$ plane as $\nu$ did before, but instead $\eta_l/4$ layers, while  $\xi \in \{1, 2, ...,\eta_l/4\}$ denotes a wire within a given $\tilde{\nu}$ (see again Fig.~\ref{fig:spectrum_l=3} for $l=3$). Note that for $l = 1$ and $l = 2$, the new unit cell coincides with the old one. In this case, the index $\tilde{\nu}$ corresponds to only one layer, making the index $\xi$ unnecessary. Consequently, the new index $\tilde{m}$ ($\tilde{\nu}$) becomes equivalent to the old index $m$ ($\nu$).

Strictly speaking, for $l>2$, the terms $H_B^{(l)}$ and $H_y^{(l)}$, given by Eqs.~(\ref{eq:Zeeman2}) and (\ref{eq:Hy}), should be rewritten using the new indices. However, since this represents only a change of variables, the terms remain equivalent in both the old and new indexing schemes. Taking into account only the $H_B^{(l)}$ and $H_y^{(l)}$ terms, we find that the system is again a stack of 2D QAH layers~\cite{Klinovaja2015}, with each layer hosting $2 l $ gapless edge states. To be more specific, each layer at $\tau = 1$ ($\tau = \bar{1}$) hosts $l$ gapless left (right) movers at the position $(\tilde{m}, \tilde{\nu}) = (1, \bar{1})$, and $l$ gapless right (left) movers at the position $(\tilde{m}, \tilde{\nu}) = (N_y, 1)$, where $N_y$ is the number of unit cells in the $y$ direction.

Next, we introduce tunneling processes in the $z$ direction. The tunneling amplitudes are assumed to depend on $\tilde{\nu}$ in the same way they before depended on $\nu$, while their periods are now independent of $l$: $t_{z \tilde{\nu} }(x) = 2 t_{z \tilde{\nu}} \cos \left(2 k_F x\right)$, and $\tilde{t}_{z \tilde{\nu} }(x) = 2 \tilde{t}_{z \tilde{\nu}} \cos \left(2 k_F x \right)$. Thus, the Hamiltonian terms for any $l$ are modified as follows:

\begin{equation}
	H_z = \sum_{n, \tilde{m}} \sum_{\tilde{\nu}, \xi, \sigma} \int d x \, t_{z \tilde{\nu}}(x) \Psi_{(n+1) \tilde{m} 1 \tilde{\nu} \xi \sigma}^{\dagger} \Psi_{n \tilde{m} \bar{1} \tilde{\nu} \xi \sigma}+\text { H.c.},
\end{equation}

\begin{equation}
	\widetilde{H}_{z} = \sum_{n, \tilde{m}} \sum_{\tilde{\nu}, \xi, \sigma} \int d x  \, \tilde{t}_{z \tilde{\nu} }(x) \Psi_{n \tilde{m} 1 \tilde{\nu} \xi \sigma}^{\dagger} \Psi_{n \tilde{m} \bar{1} \tilde{\nu} \xi \sigma}+\text { H.c.},
\end{equation}
where $\Psi_{n \tilde{m} \tau \tilde{\nu} \xi \sigma}(x)$ is the annihilation operator of an electron with spin $\sigma \in \{1, \bar{1} \}$ at the position $x$ of the wire $(\tau, \tilde{\nu}, \xi)$ in the unit cell $(n, \tilde{m})$.  Following the same parameter hierarchy  $t_{z 1}, \tilde{t}_{z \bar{1} } \gg t_{z \bar{1}}, \tilde{t}_{z 1}$, we find that all states in the bulk and on the surfaces of the sample are fully gapped out, while $2l$ states localized to two hinges of the sample are left gapless: $l$ right movers at $(n, \tilde{m}, \tilde{\nu}) = (1, N_y, 1)$ and $l$ left movers at $(n, \tilde{m}, \tilde{\nu}) = (N_z, N_y, 1)$, see the red ovals in Fig.~\ref{fig:spectrum_l=3} for the case $l = 3$.

\end{document}